\documentclass[11pt]{iopart2}

\usepackage{epsfig,amssymb,amsmath,bm,graphicx}

\def\spose#1{\hbox to 0pt{#1\hss}}
\def\lesssim{\mathrel{\spose{\lower 3pt\hbox{$\mathchar"218$}}
 \raise 2.0pt\hbox{$\mathchar"13C$}}}
\def\gtrsim{\mathrel{\spose{\lower 3pt\hbox{$\mathchar"218$}}
 \raise 2.0pt\hbox{$\mathchar"13E$}}}

\def\<{\langle}
\def\>{\rangle}

\usepackage[english]{babel}
\usepackage[T1]{fontenc}
\usepackage[utf8]{inputenc}

\begin{document}

\title{
Disordered Ising model with correlated frustration
}

\author{Angelo Giorgio Cavaliere$^{1}$ and
Andrea Pelissetto$^{1,2}$ }

\address{$^1$
Dip. Fisica dell'Universit\`a di Roma ``La Sapienza", Piazzale Aldo Moro 2,\\
I-00185 Roma, Italy}

\address{$^2$
INFN, Sezione di Roma 1, Piazzale Aldo Moro 2,\\
I-00185 Roma, Italy}

\ead{
Angelogiorgio.Cavaliere@uniroma1.it,
Andrea.Pelissetto@roma1.infn.it}

\begin{abstract}
We consider the $\pm J$ Ising model on a cubic lattice with a gauge-invariant disorder distribution. Disorder depends on a parameter $\beta_G$ that plays the role of a chemical potential for the amount of frustration. 
We study the model at a specific value of the disorder parameter $\beta_G$,
where frustration shows long-range correlations. We characterize the 
universality class, obtaining accurate estimates of the critical exponents:
$\nu = 0.655(15)$ and $\eta_q = 1.05(5)$, where $\eta_q$ is the overlap
susceptibility exponent.
\end{abstract}

\maketitle

\section{Introduction}

The $\pm J$ Ising model is a paradigmatic model
to study the effects of quenched random disorder and
frustration. It is defined by the lattice Hamiltonian
\begin{equation}
{\cal H} = - \sum_{\langle xy \rangle} J_{xy} \sigma_x \sigma_y,
\label{HamJ}
\end{equation}
where $\sigma_x=\pm 1$ and the sum is over the nearest-neighbor sites of a 
lattice. In the following we will focus on the three-dimensional system defined on a simple
cubic lattice. Usually, the exchange interactions $J_{xy}$ are uncorrelated
quenched random variables, taking values $\pm 1$ with probability distribution
\begin{equation}
P(J_{xy}) = p \delta(J_{xy} - 1) + (1-p) \delta(J_{xy} + 1). 
\label{prob-EA}
\end{equation}
For $p=1$ we recover the standard ferromagnetic
Ising model, while for $p=1/2$ we obtain the bimodal 
Edwards-Anderson (EA) spin-glass model.  
The $\pm J$ Ising model is a simplified model \cite{EA-75} for
some disordered magnetic materials,
such as \cite{IATKST-86,GSNLAI-91,NN-07} Fe$_{1-x}$Mn$_x$TiO$_3$ and
Eu$_{1-x}$Ba$_x$MnO$_3$
that have a low-temperature glassy phase.

The behavior of the three-dimensional $\pm J$ Ising model is 
well known. In this model, disorder is a relevant perturbation of the 
pure Ising fixed point: the introduction of any amount of disorder in the 
couplings changes the universality class \cite{Aharony-76,PV-02-review}.
The $T$-$p$ phase diagram is
characterized by a high-temperature paramagnetic phase separated by two
different low-temperature phases by a transition line. If we only consider the 
case $p > 1/2$ (the phase diagram is symmetric under $p\to 1-p$), 
we have a ferromagnetic transition for $p > p^*$ and a glassy transition 
in the opposite case.  In the first case, 
the transition belongs to the randomly-dilute Ising (RDI)
universality class \cite{HPPV-07-pmj} characterized by
the magnetic critical exponents
\cite{BFMMPR-98,PV-02,HPPV-07} $\nu_{f}=0.683(2)$ (quite different 
from the Ising value $\nu_I \approx 0.630$ \cite{PV-02-review})
and $\eta_{f}=0.036(1)$. 
On the other hand, for any $1-p^* < p < p^*$, the transition
belongs to the same universality class as that of the bimodal Ising
spin glass model at $p=1/2$, with \cite{KKY-06,HPV-08,Janus-13,LPP-16}
$\nu_{SG} =  2.56(4)$ and $\eta_{q,SG} = -0.390(4)$. 
The point $p = p^* = 0.76820(4) $ is a multicritical point 
\cite{Nishimori-81,Nishimori-86,Nishimori-book,HPPV-07-multicr}.
The renormalization-group (RG) dimensions $y_1$ and $y_2$ of the relevant
operators that control the RG flow close to it 
are \cite{HPPV-07-multicr}
$y_1 = 1.02(5)$, $y_2 = 0.61(2)$. Correspondingly, 
the temperature and crossover exponents are
$\nu=1/y_2=1.64(5)$ and $\phi=y_2/y_1 = 1.67(10)$, respectively.  

In this paper we consider again the $\pm J$ model with Hamiltonian 
(\ref{HamJ}), but we use a different probability distribution. We do not
directly fix the distribution of the couplings, but rather we 
specify the amount of frustration present in the system. Moreover, we use
a gauge-invariant probability distribution: two sets of bonds that correspond
to the same frustration distribution are given the same probability.
Frustration is quantified by using the product of the couplings 
$J_{xy}$ corresponding 
to the links along an elementary lattice square (plaquette) and the 
corresponding probability distribution is nothing but the Boltzmann-Gibbs 
probability for the 
lattice ${\mathbb Z}_2$ gauge theory \cite{BDI-75,Kogut-79} at inverse 
temperature $\beta_G$. In this context the parameter $\beta_G$ plays the 
role of a chemical potential for the frustrated lattice plaquettes.
Since the probability distribution depends on the values of the couplings 
along the different plaquettes, the bond variables on different links are 
necessarily correlated. If these correlations are very short-ranged 
(this is the case for $|\beta_G|$ small), we expect the behavior to be 
analogous to that of the usual EA model.\footnote{A different 
correlated distribution was considered in Ref.~\cite{PTPV-06}: the critical
behavior at the transition was the same as that in the EA model with 
coupling distribution (\ref{prob-EA}).}  On the other hand, as $\beta_G$
increases, correlations become strong and we expect a different phase diagram.
Here, we focus on a specific value $\beta_{G,c}$ 
of $\beta_G$ that corresponds to a phase 
transition point of the exchange coupling distribution. At $\beta_{G,c}$ 
frustration shows long-range power-law correlations. In the
temperature-$\beta_G$ plane, close to this point, the system shows a 
multicritical behavior. We study it in detail, identifying the universal 
features of the new universality class. We also report some preliminary
results for the behavior for $\beta_G > \beta_{G,c}$. In this range of values
of $\beta_G$, the model apparently behaves as the pure ferromagnetic Ising
model. These results, if confirmed by simulations on larger lattices, 
imply that the ferromagnetic transition in pure systems is stable with 
respect to the introduction of a small amount of correlated disorder. 
This is at variance with what happens for uncorrelated disorder, i.e.,
in the case one uses distribution (\ref{prob-EA}). In the latter case,
the introduction of any amount of antiferromagnetic bonds changes the 
universality class of the system \cite{Aharony-76,PV-02-review}.

The paper is organized as follows. In section \ref{sec2} we present
the model, while in section \ref{sec3} we define the quantities
we measure in the simulation.  In section \ref{sec4} we present our
Monte Carlo results at the transition of the gauge model, while 
some preliminary results in the exchange-coupling low-temperature phase 
are presented in section \ref{sec5}. We draw our conclusions in 
section \ref{sec6}. Some computational details are reported in
the Appendix.

\section{The model} \label{sec2}

We consider the $\pm J$ Ising model with quenched correlated disorder.
The model is defined on a cubic lattice of linear size $L$ in all 
directions, with Hamiltonian (\ref{HamJ}).
In the usual EA model, the bond couplings on different links are 
uncorrelated and distributed with probability distribution 
(\ref{prob-EA}).
The presence of negative couplings makes the EA model frustrated. 
A simple measure of frustration is provided by the product $\Pi_P$ of the 
couplings $J_{xy}$ along an elementary lattice square (plaquette) $P$:
\begin{equation}
   \Pi_p = J_{\ell_1} J_{\ell_2} J_{\ell_3} J_{\ell_4},
\end{equation}
where $\ell_1$, $\ell_2$, $\ell_3$, and $\ell_4$ are the four links belonging
to the plaquette $P$.
If $\Pi_P < 0$, the plaquette is frustrated: it is not possible
to minimize the local energy on each link of the plaquette, i.e., it is 
not possible that $J_{xy} \sigma_x \sigma_y > 0$ on all four links 
$\ell_1,\ldots,\ell_4$.

In this work we consider a different probability distribution for 
the bond couplings that allows us to tune directly 
the amount of frustration in the system. We consider 
\begin{equation}
P(\{J_{xy}\}) = {1\over Z_G} e^{-\beta_G H_G} \qquad 
   H_G = - \sum_P \Pi_P,
\label{H-gauge}
\end{equation}
where the sum is over all lattice plaquettes and $Z_G$ is a normalizing 
factor.  The parameter $\beta_G$ plays the role of a chemical potential for 
the frustrated plaquettes. For $\beta_G\to \infty$, we obtain an 
unfrustrated model, while maximal frustration is obtained in the opposite limit 
$\beta_G \to -\infty$. For $\beta_G = 0$ the model is equivalent to the 
EA one with $p=1/2$.  Note that the probability distribution 
can also be written as 
\begin{equation}
P(\{J_{xy}\}) = {1\over Z_G} \prod_P 
  \left[e^{\beta_G} \delta(\Pi_P - 1) + e^{-\beta_G} \delta(\Pi_P + 1)\right],
\end{equation}
that explicitly shows the role of $\beta_G$ in controlling frustration.

The model defined by Eq.~(\ref{H-gauge}) is the well-known lattice 
${\mathbb Z}_2$ gauge theory \cite{BDI-75,Kogut-79}, 
which is invariant under the 
${\mathbb Z}_2$ gauge transformations 
\begin{equation}
  J_{xy} \to J_{xy}' = \epsilon_x J_{xy} \epsilon_y,
\end{equation}
for any arbitrary site-dependent $\epsilon_x = \pm 1$. Note that also 
Hamiltonian (\ref{HamJ}) is gauge invariant, provided 
the Ising spins are transformed according to
\begin{equation}
\sigma_x \to \sigma_x '= \epsilon_x \sigma_x.
\end{equation}
Using duality \cite{Kogut-79,Savit-80}, for 
positive values of the coupling $\beta_G$, 
it is possible to map the gauge theory (\ref{H-gauge}) 
onto the pure Ising model. Therefore, for $\beta_G > 0$,
the gauge model has two 
different phases separated by a continuous phase transition 
located at $\beta_{G,c}$, which can be related to the transition inverse 
temperature $\beta_{I,c}$ in the pure Ising model by 
\begin{equation}
   \beta_{G,c} = \frac{1}{2}\ln\coth\beta_{I,c}.
\end{equation}
The pure Ising critical temperature is known
with high-precision \cite{Hasenbusch-10,FXL-18,KPSDV-16}. 
Using \cite{FXL-18} $\beta_{I,c} = 0.221654626(5)$, 
we obtain $\beta_{G,c} = 0.761413292(11)$. It is interesting to note 
that the distribution (\ref{H-gauge}) is also well defined for negative 
$\beta_G$.
In this case the behavior of the gauge model is not related to that 
of the pure Ising model.

\section{Observables} \label{sec3}

Since the theory is gauge invariant, we should only consider 
gauge invariant quantities. As usual in studies of glassy systems, we 
consider observables defined in terms of the overlap variable $q_x$ 
\begin{equation}
q_x = \sigma^{(1)}_{x}\sigma^{(2)}_{x},
\end{equation}
where the indices in parentheses refer to two replicas of the system with the 
same bond couplings.  Note that the overlap variable $q_x$ is 
gauge invariant and therefore 
thermal averages of any function of $q_x$ are identical for 
coupling distributions that differ by a gauge transformation. 
We define the Binder cumulant 
\begin{equation}
U_{q,4} = 
\frac{\bigl[ \langle Q^4 \rangle\bigr]}{\bigl[\langle Q^2 \rangle\bigr]^2}
\qquad  Q = \sum_x q_x,
\label{eq:u4q}
\end{equation}
where the angular and square brackets refer
to the thermal average and to the quenched average over the bond 
distribution, respectively. The Binder parameter $U_{q,4}$
plays an important role in our analysis as it 
is renormalization-group (RG) invariant. 
We also consider the overlap correlation function
\begin{equation}
 G_{q}(x) = {1\over V} \sum_y [\langle q_y q_{x+y} \rangle ] 
          = {1\over V} \sum_y [\langle \sigma_y \sigma_{x+y}\rangle^2]
\qquad 
 \tilde{G}_q(k) = \sum_x e^{ik\cdot x} G_q(x) ,
\end{equation}
($V = L^3$ is the volume),
the corresponding susceptibility 
$\chi_q = \tilde{G}_q(0) = [\langle Q^2\rangle]/V$ and 
correlation length
\begin{equation}
\xi^2_{q} = \frac{\chi_q-\chi_1}{4\sin^2(\pi/L)\chi_1},
\label{eq:xiq}
\end{equation}
where 
\begin{equation}
\chi_1 = \frac{1}{3} \Bigl[ \tilde{G}_q(2\pi/L, 0, 0) + 
   \tilde{G}_q(0, 2\pi/L, 0) +  \tilde{G}_q(0, 0, 2\pi/L)\Bigr] .
\end{equation}
Note that $R_\xi = \xi_q/L$ is RG invariant.

\section{Multicritical behavior close to the critical point $\beta_G$}
\label{sec4}

The behavior of the disordered system depends on both $\beta_G$ and $\beta$.
Since the gauge model has a transition for $\beta_G = \beta_{G,c}$,
we expect at least three different universality classes for the 
behavior of the spin system, one for $\beta_G > \beta_{G,c}$, one 
for $\beta_{G} < \beta_{G,c}$ and one for $\beta_{G} = \beta_{G,c}$. 
In particular, if the spin 
system has a transition for $\beta = \beta_c$ when $\beta_G = \beta_{G,c}$, 
the point $(\beta_{G,c},\beta_c)$ is a
multicritical point in the parameter space $(\beta_G,\beta)$.
The relevant scaling variables are two 
linear scaling fields $t_1$ and $t_2$, that are linear combinations 
of $\beta_G$ and $\beta$. However, the transition in the gauge model 
is unrelated to the value of $\beta$ and hence one scaling field 
should simply be $t_1 = \beta_G - \beta_{G,c}$. The second one should be
instead 
$t_2 = \beta - \beta_c + a (\beta_G - \beta_{G,c})$, where $a$ is an 
appropriate constant.
In a neighborhood of the  multicritical point, any RG invariant 
quantity $R$ behaves as 
\begin{equation}
R = f_R(t_1 L^{1/\nu_I},t_2 L^{1/\nu}) ,
\label{R-RG}
\end{equation}
where $\nu_I$ is the pure Ising exponent 
($\nu_I = 0.629971(4)$ in the conformal-boostrap approach 
\cite{KPSDV-16} and 
$\nu_I = 0.63002(10)$, $0.62991(9)$ numerically \cite{Hasenbusch-10,FXL-18})
and $\nu$ is a new critical exponent characterizing the universality class.

\begin{figure*}[!tbph]
\begin{center}
\includegraphics[width=0.5\textwidth, keepaspectratio]{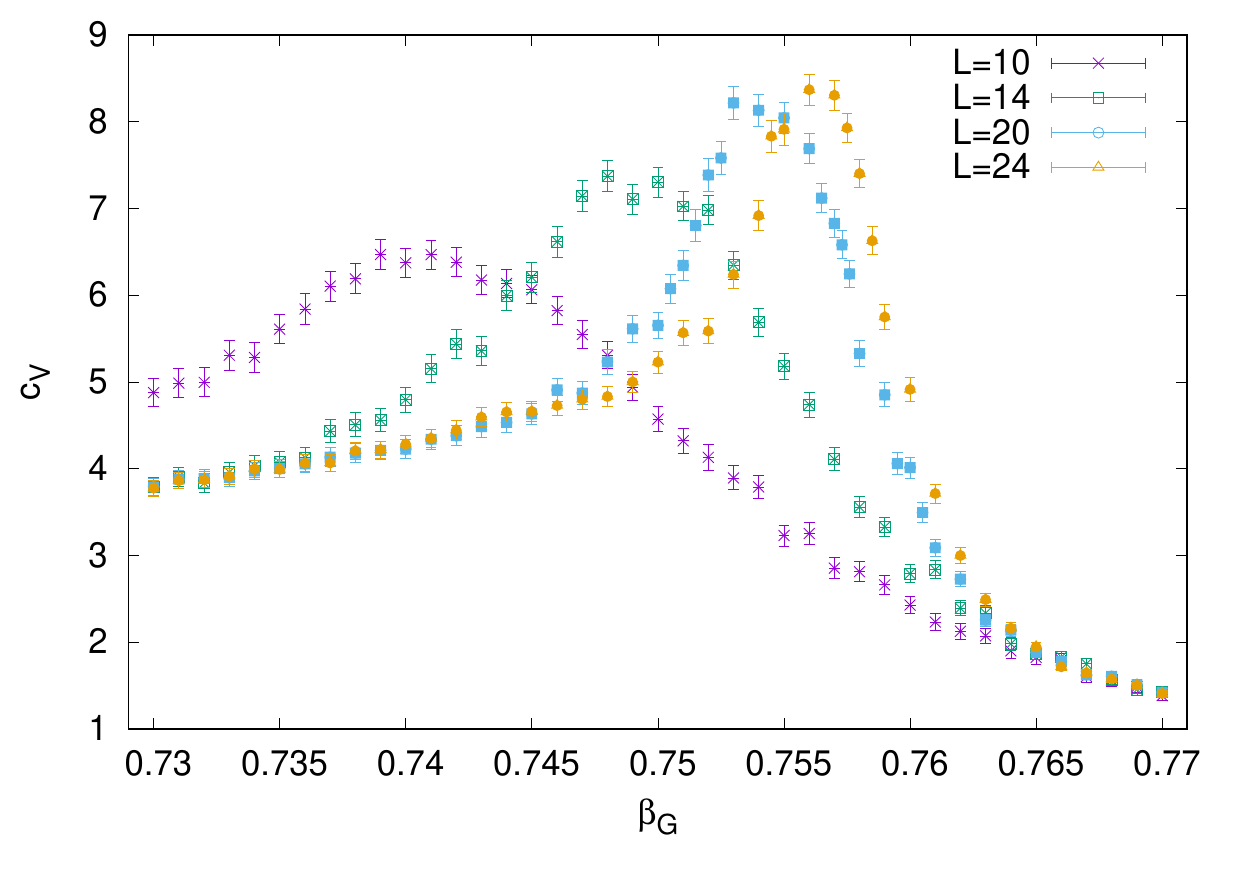}
\end{center}
\caption{Specific heat $C_v$ of the gauge 
${\mathbb Z}_2$ model as a function of $\beta_G$.  
}
\label{specific-heat}
\end{figure*}

To determine the critical behavior of the system, we will study two different
cases. First, we perform simulations setting $\beta_G = \beta_{G,c}$. In this case,
Eq.~(\ref{R-RG}) becomes
\begin{equation}
R = f_R(0,X) \qquad X \equiv (\beta - \beta_c) L^{1/\nu}.
\label{R-RG2}
\end{equation}
Second,
we consider the behavior at the finite-size pseudocritical transition of 
the gauge model, defined by considering the maximum of the 
specific heat $C_v$, 
\begin{equation} 
C_v = {\beta_G^2\over V} \sum_{P, Q} 
   \left[ \langle \Pi_{P}  \Pi_Q \rangle - \langle \Pi_{P}\rangle
   \langle  \Pi_Q \rangle \right],
\end{equation}
where the sum runs over all pairs of plaquettes $P$ and $Q$.
In figure~\ref{specific-heat} we report the specific heat as a function 
of $\beta_G$. For each $L$ in the range $10\le L \le 30$, 
we determine the position $\beta_G(L)$ of the maximum,
by performing a quadratic fit close to the maximum. 
The results are then fitted to 
\begin{equation}
\beta_G(L) = \beta_{G,c} + X_{\rm max} L^{-1/\nu_I},
\label{betaGL-def}
\end{equation}
to obtain the constant $X_{\rm max}$. Using \cite{Hasenbusch-10}
$\nu_I = 0.63002$, we obtain 
$X_{\rm max} = -0.8080(53)$. The quality of the fit is good: The sum of the 
residuals ($\chi^2$) normalized to the number of degrees of freedom (DOF) is 
approximately 0.4.
Therefore, we perform a second set of simulations varying 
$\beta_G$ with $L$ as in Eq.~(\ref{betaGL-def}), using $X_{\rm max} = -0.8080$.
Note that, for this set of runs, we have
\begin{equation}
t_2 = \beta - \beta_c + a X_{\rm max} L^{-1/\nu_I},
\end{equation}
and, therefore, we can simply set $t_2 = \beta - \beta_c$, neglecting scaling 
corrections. The comparison of the results of the simulations at fixed 
$\beta_G = \beta_{G,c}$ and at $\beta_G(L)$ will provide an important check
for the final results. Indeed, we expect quite different $L$ corrections in the
two cases, since, for finite $L$, we are sampling a somewhat different set of
configurations. 

Beside RG invariant quantities, we will also consider the susceptibility 
$\chi_q$, which scales close to the multicritical point as  
\begin{equation}
\chi_q(\beta,L) = L^{2-\eta_q} u_\chi(t_1,t_2) 
    f_\chi(t_1 L^{1/\nu_I},t_2 L^{1/\nu}) ,
\label{chi-RG}
\end{equation}
where $\eta_q$ is a universal exponent and $u_\chi$ is a nonlinear 
scaling field. 

\subsection{Behavior of the overlap observables 
at the gauge critical point}

\begin{figure*}[!tbh]
\begin{tabular}{cc}
\includegraphics[width=0.5\textwidth,keepaspectratio]{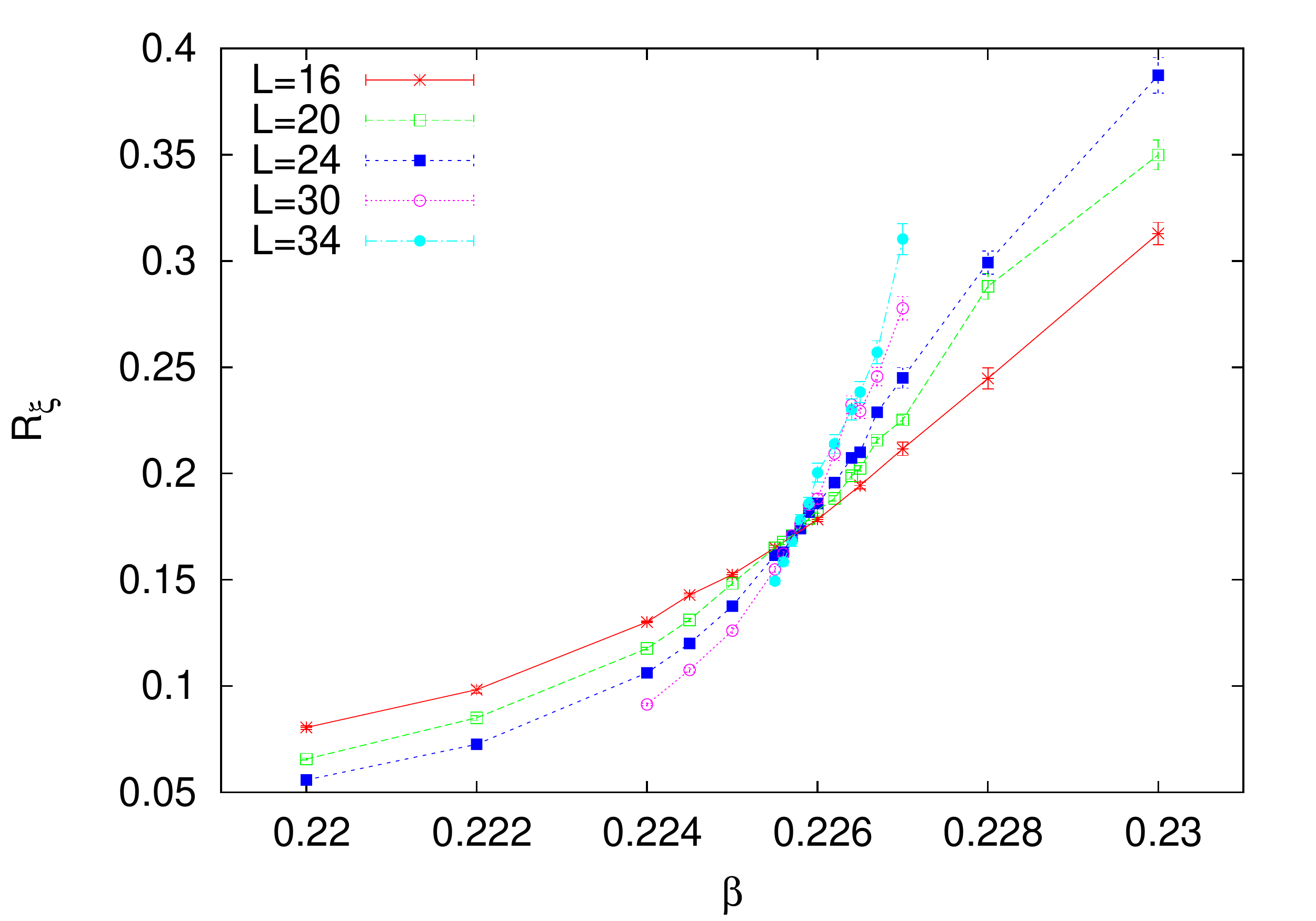} &
\includegraphics[width=0.5\textwidth,keepaspectratio]{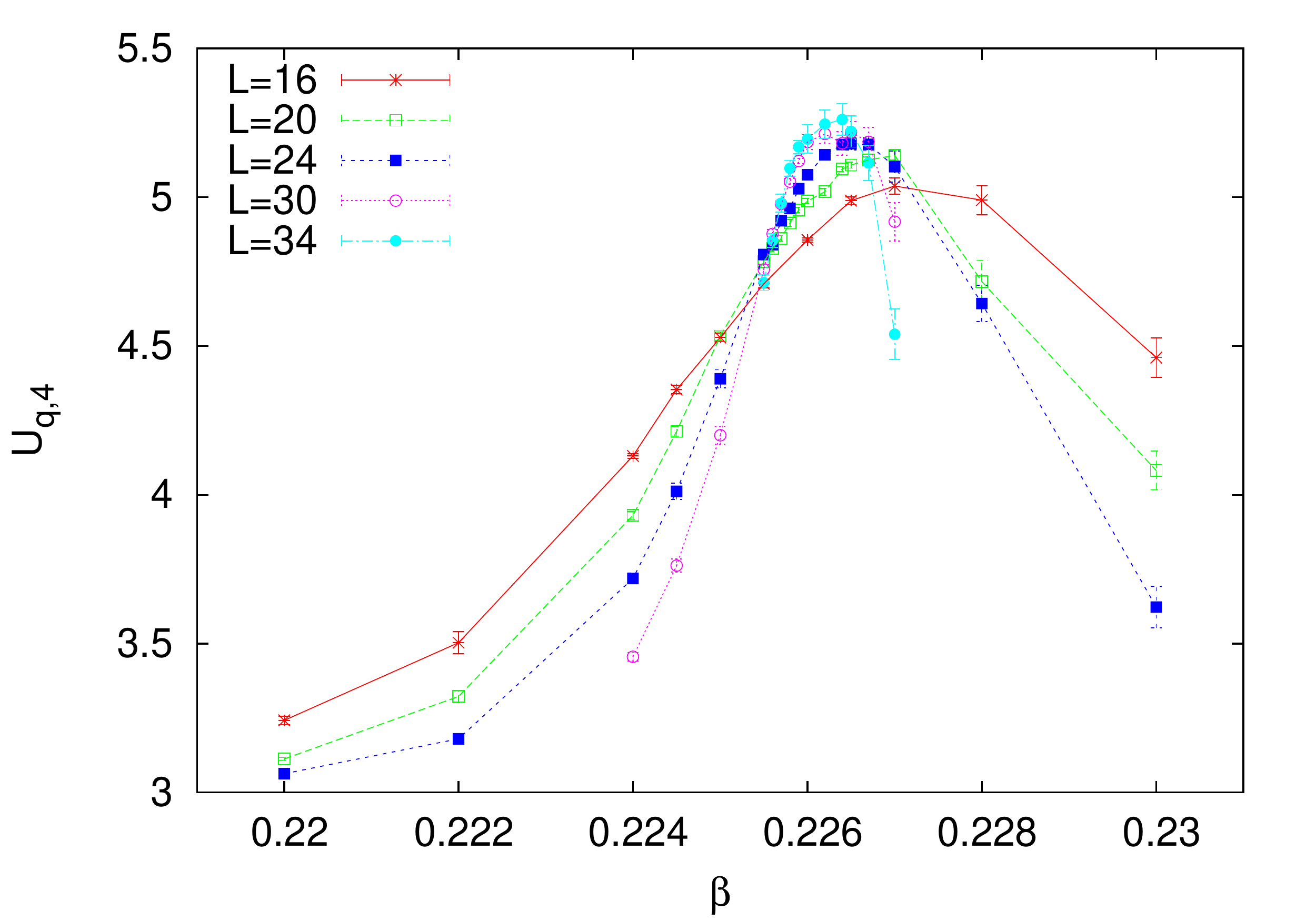} \\
\end{tabular}
\caption{Plot of $R_\xi = \xi_q/L$ (left) and of 
$U_{q,4}$ (right) versus $\beta$ for several values of $L$. 
All results are obtained for $\beta_G=\beta_{G,c}$. 
Lines are only meant to guide the eye.
}
\label{fig:case-a-overlap}
\end{figure*}

We have performed simulations at $\beta_{G,c}$ for several values of 
$L$ in the range $16\le L \le 34$. Computational details
are reported in the Appendix. Results for $R_\xi = \xi_q/L$ 
and the Binder parameter $U_{q,4}$ are reported in
figure~\ref{fig:case-a-overlap}. Data show an intersection for
$\beta \simeq 0.226$, that gives us a first estimate of the critical 
temperature. To obtain more precise estimates, we use 
Eq.~(\ref{R-RG2}). Assuming that $X$ is small, we can expand 
the scaling function in powers of $X$. Then,
we fit any RG invariant quantity $R$ to 
\begin{equation}
   R(\beta,L) = R^* + \sum_{k=1}^n a_k X^k \qquad 
X \equiv (\beta - \beta_c) L^{1/\nu},
\label{fits-R}
\end{equation}
where $R^* = f_R(0,0)$ is a universal quantity.
Here $n$ is the order of the polynomial in $X$, which is tuned by looking at 
the quality of the fit.
We have not included scaling corrections. To estimate their 
role, we repeat the fits for $L_{\rm min} = 16,20,24$, each time only 
including data corresponding to $L\ge L_{\rm min}$.

\begin{table}[tbph]
\begin{center}
\begin{tabular}{rcccccc }
\hline\hline
 & $n$ & $L_{\rm min}$ & $\beta_{c}$&$R^*$ &$\nu$ &$\chi^2$/DOF \\
			\hline
$R_\xi$   & 5 & 16 & 0.225753(12) & 0.1726(6) & 0.681(9) & 1.4 \\
          & 6 & 16 & 0.225755(12) & 0.1729(6) & 0.676(10)& 1.3 \\
          & 5 & 20 & 0.225767(14) & 0.1735(8) & 0.678(14)& 1.5 \\
          & 6 & 20 & 0.225766(14) & 0.1736(8) & 0.664(15)& 1.4 \\
          & 5 & 24 & 0.225756(23) & 0.1728(14)& 0.678(27)& 1.4 \\
          & 6 & 24 & 0.225756(23) & 0.1729(14)& 0.654(28)& 1.3 \\
$U_{q,4}$ & 7 & 16 & 0.225364(12) & 4.686(7)  & 0.653(10)& 5.8 \\
          & 8 & 16 & 0.225369(13) & 4.687(7)  & 0.653(10)& 5.9 \\
          & 7 & 20 & 0.225446(15) & 4.748(10) & 0.631(16)& 4.2 \\
          & 8 & 20 & 0.225448(15) & 4.749(10) & 0.634(16)& 4.3 \\
          & 7 & 24 & 0.225536(23) & 4.812(20) & 0.631(33)& 1.9 \\
          & 8 & 24 & 0.225537(23) & 4.817(20) & 0.614(33)& 1.9 \\
\hline\hline
\end{tabular}
\end{center}
\caption{Results (data at $\beta_G = \beta_{G,c}$) of the fits to 
Eq.~(\ref{fits-R}): the order $n$ of the expansion is
given in the second column.
In each fit we only include the data satisfying $L\ge L_{\rm min}$.
In the last column we report the sum of the residuals divided by the 
number of degrees of freedom (DOF) of the fit.
}
\label{table:results-fits}
\end{table}

We first consider $\xi_q/L$. For each $L_{\rm min}$ we increase
systematically the order $n$ till results and the $\chi^2$ per degree of freedom
are stable. For all values of $L_{\rm min}$ this is obtained by taking
$n\approx 5$-6, see Table~\ref{table:results-fits}. 
The quality of the fits is reasonable and results are stable with 
respect to $L_{\rm min}$, indicating that scaling corrections are 
small, compared to the statistical errors. 
From this analysis we would estimate $\beta_c = 0.22576(2)$ and 
$\nu = 0.67(3)$.

The estimates obtained analyzing $U_{q,4}$ are worse. 
Independently of $n$, the quality of the fits is poor, 
see Table~\ref{table:results-fits}.
Moreover, results for different values of $L_{\rm \min}$ show a systematic 
drift, a clear signal of the presence of systematic errors due to the neglected 
scaling corrections. To obtain more reliable estimates, it is therefore
necessary to include scaling corrections. We have thus performed a
combined analysis of $R_\xi$ and $U_{q,4}$ in which 
$R_\xi$ has been fitted to (\ref{fits-R}), while 
$U_{q,4}$ has been fitted to 
\begin{equation}
   R(\beta,L) = R^* + \sum_{k=1}^n a_k X^k + 
     L^{-\omega} \sum_{k=0}^m b_k X^k
     \qquad 
X \equiv (\beta - \beta_c) L^{1/\nu}.
\label{fits-R2}
\end{equation}
Fits are stable only if we take $L_{\rm min} = 16$, that is, if we include all
data. Using $n = 6$ for $R_\xi$
and $n = 8$, $m=2$ for $U_{q,4}$ we obtain a good-quality fit: 
$\chi^2$/DOF = 1.14. Correspondingly, we obtain the estimates
\begin{equation}
\beta_{c}=0.225754(11) \qquad \nu = 0.666(8) \qquad \omega = 1.18(25).
\label{final-est1}
\end{equation}
Moreover, we have $R_\xi^* = 0.1728(6)$ and $U_{q,4}^* = 5.20(6)$. 
The estimates of $\beta_c$ and $\nu$ 
are perfectly consistent with those obtained before from
the analysis of $R_\xi$ alone. To verify the reliability of the errors we have
performed a second fit. Both observables are fitted to (\ref{fits-R2}), with 
$n=6$, $m=2$ for $R_\xi$ and $n=8$, $m=2$ for $U_{q,4}$. We obtain a 
sligthly better $\chi^2$/DOF, 1.10, and 
\begin{equation}
\beta_{c}=0.225823(37) \qquad \nu = 0.637(14) \qquad \omega = 0.83(27).
\label{final-est2}
\end{equation}
Moreover, we have $R_\xi^* = 0.182(7)$ and $U_{q,4}^* = 5.43(20)$. The new
estimates are essentially consistent with the results reported in 
(\ref{final-est1}), differences
being of the order of the sum of the statstical errors. 

In fit (\ref{fits-R2}) we have included nonanalytic corrections related to the 
leading irrelevant operator. Another source of scaling corrections is the 
analytic background. For $R_\xi$ and $U_{q,4}$ 
such corrections decay as $L^{\eta_q - 2}$, where $\eta_q$ is the 
susceptibility exponent. Usually, such an exponent is small and analytic
corrections are negligible. In the present case, however, as we discuss 
in Sec.~\ref{sec:etaq}, $\eta_q \approx 1$ and thus they are as important as 
the nonanalytic ones. We have therefore performed fits to 
\begin{equation}
   R(\beta,L) = R^* + \sum_{k=1}^n a_k X^k + 
     L^{\eta_q - 2} \sum_{k=0}^m b_k (\beta - \beta_c)^k
     \qquad 
X \equiv (\beta - \beta_c) L^{1/\nu},
\label{fits-R3}
\end{equation}
where we have only considered the analytic terms in $(\beta - \beta_c)$
(in principle, the coefficient function in front of $L^{\eta_q - 2}$ 
has an expansion in powers of both $(\beta - \beta_c)$ and $X$).
We fix $\eta_q = 1.05$, see Sec.~\ref{sec:etaq}. A combined analysis 
of the two observables, using $n = 6$, $m=1$ for $R_\xi$ and 
$n = 8$, $m=1$ for $U_{q,4}$ gives $\chi^2$/DOF = 1.25 and 
\begin{equation}
\beta_{c}=0.225742(19) \qquad \nu = 0.657(11) ,
\label{final-est3}
\end{equation}
$R_\xi^* = 0.172(3)$ and $U_{q,4}^* = 5.26(3)$. These results are fully 
consistent with the previous ones. Collecting all results we end up with the
estimates
\begin{equation}
\beta_{c}=0.22577(5) \qquad \nu = 0.655(15) \qquad \omega = 1.0(3).
\label{final-est}
\end{equation}
It is quite interesting to observe that 
the critical point position is close to 
that of the pure Ising model ($\beta_{C,I} \simeq 0.22165$). 
On the other hand, $R_\xi^*$ and $U_{q,4}^*$ are distinctly different 
since $R_{\xi}^* \approx 0.4$ and $U_{q,4} \approx 2.4$ 
in the Ising model. 

\begin{figure}[!tbph]
\begin{tabular}{cc}
\includegraphics[width=0.5\textwidth,keepaspectratio]{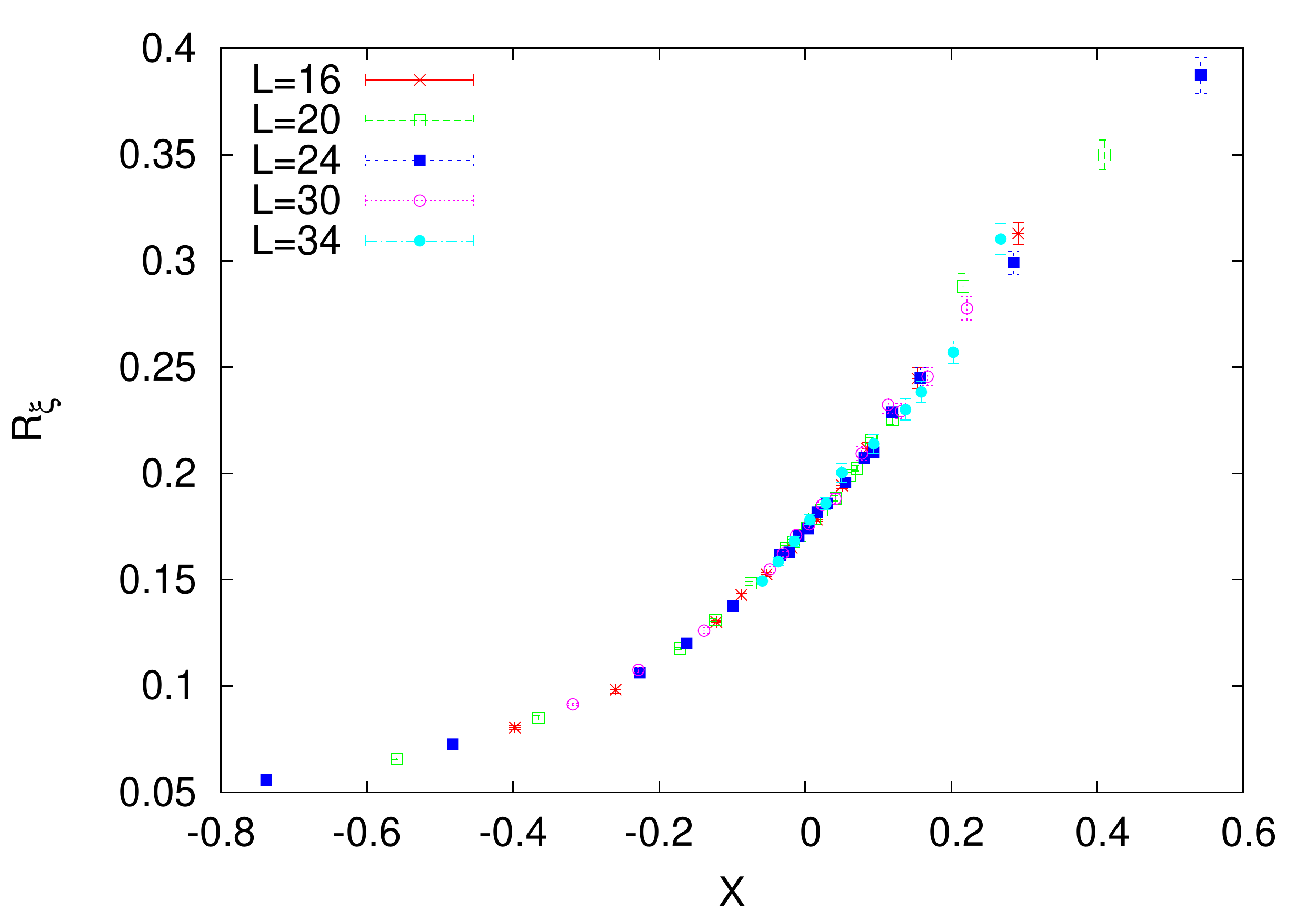} &
\includegraphics[width=0.5\textwidth,keepaspectratio]{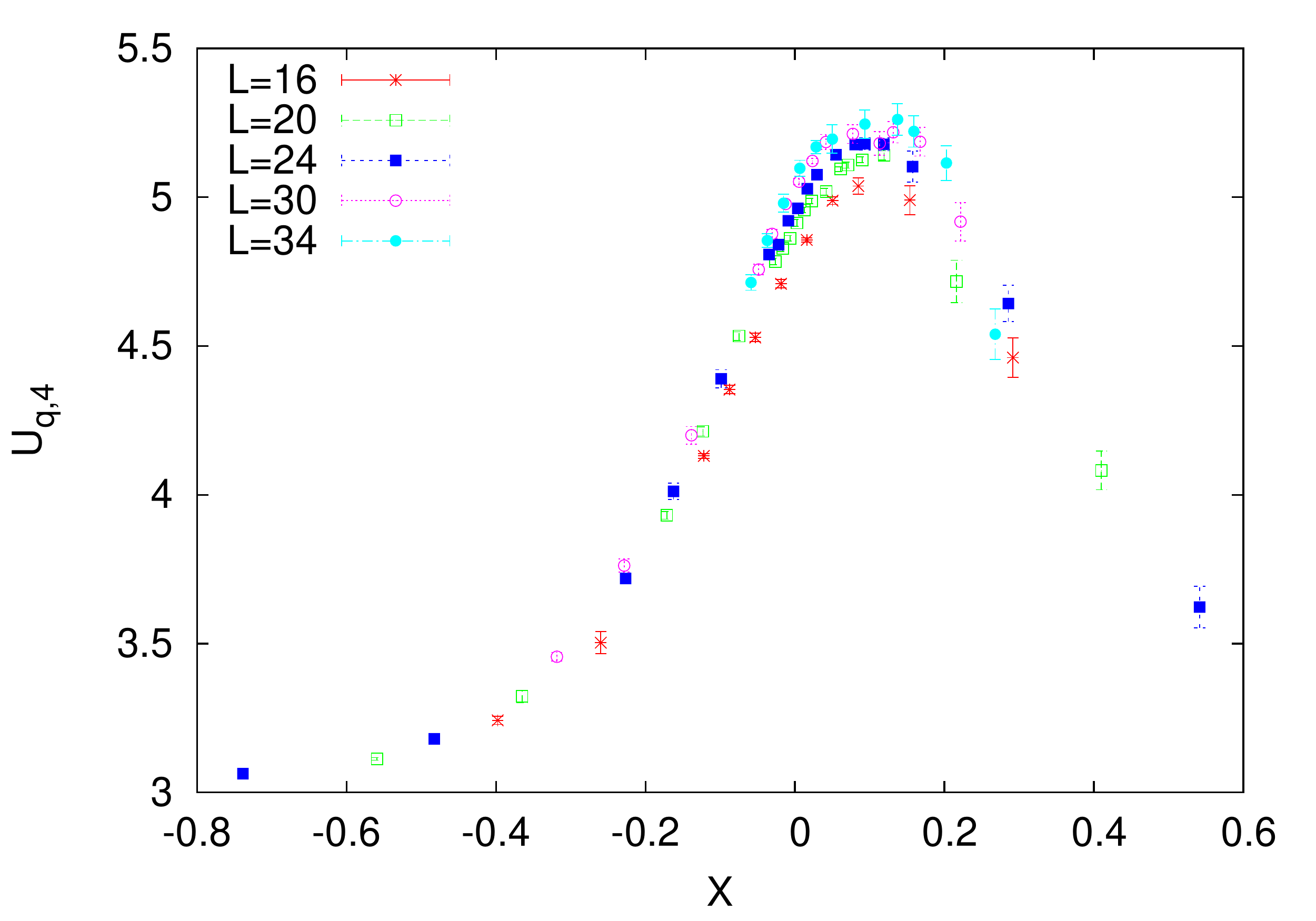} \\
\end{tabular}
\caption{Results for $R_\xi = \xi_q/L$ (left) and $U_{q,4}$ (right) 
as a function of $X = (\beta - \beta_c) L^{1/\nu}$ for several values of $L$. 
All results for $\beta_G=\beta_{G,c}$. We set 
$\beta_c = 0.22577$ and $\nu = 0.655$. 
}
\label{fig:case-a-collapse}
\end{figure}

In figure~\ref{fig:case-a-collapse}  we report the estimates of $R_\xi$
and $U_{q,4}$ as a function of 
$X = (\beta - \beta_c) L^{1/\nu}$. For $R_\xi$ we observe a very good scaling:
data are clearly consistent with the predicted scaling behavior. On the 
other hand, the Binder parameter shows 
scaling corrections, that appear to be particularly large
where the scaling function has a maximum, for $X \approx 0.1$. 

\subsection{Behavior of the overlap observables 
at the finite-size pseudocritical transition}

\begin{figure*}[tbph]
\includegraphics[width=0.5\textwidth,keepaspectratio]{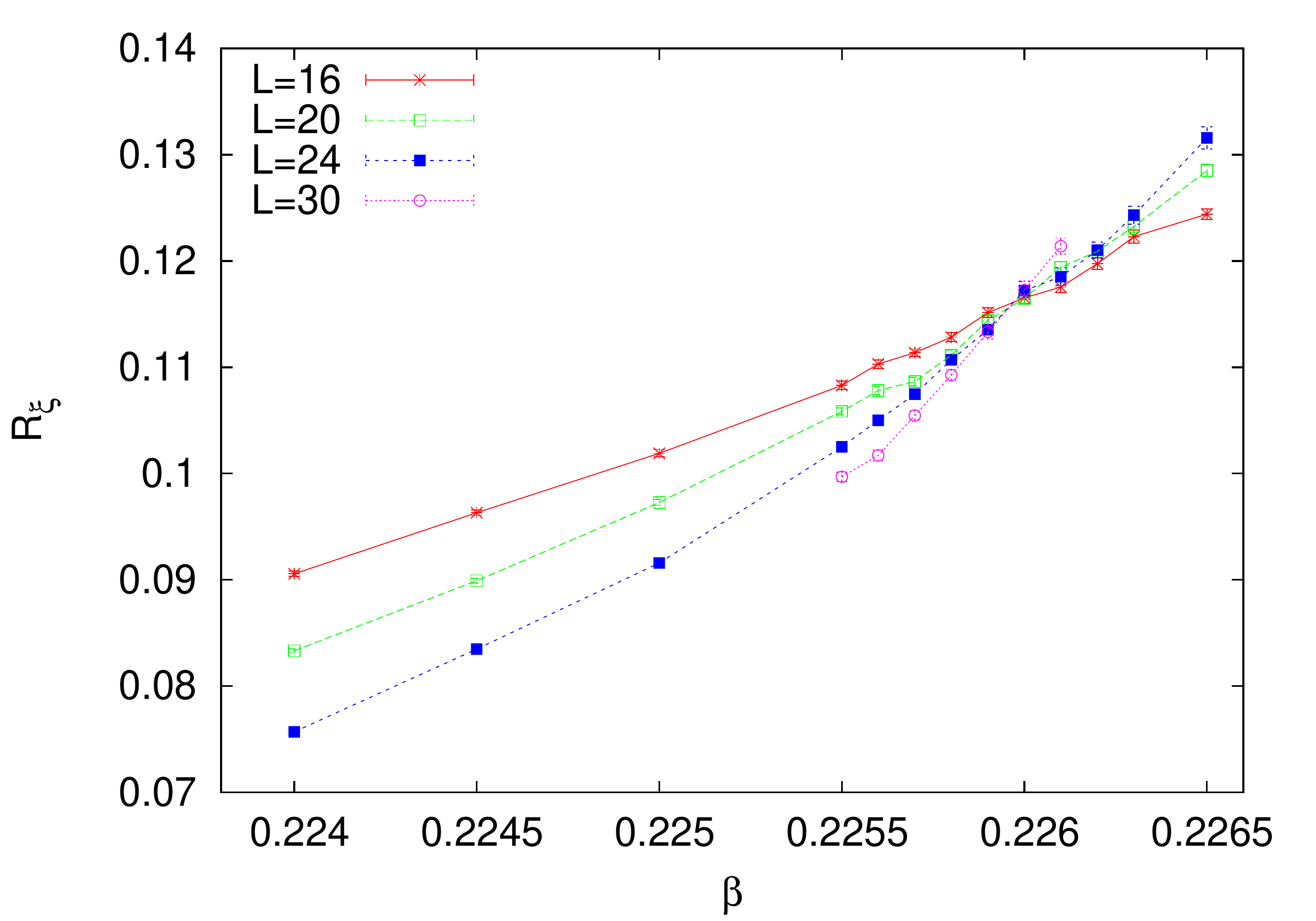}
\includegraphics[width=0.5\textwidth,keepaspectratio]{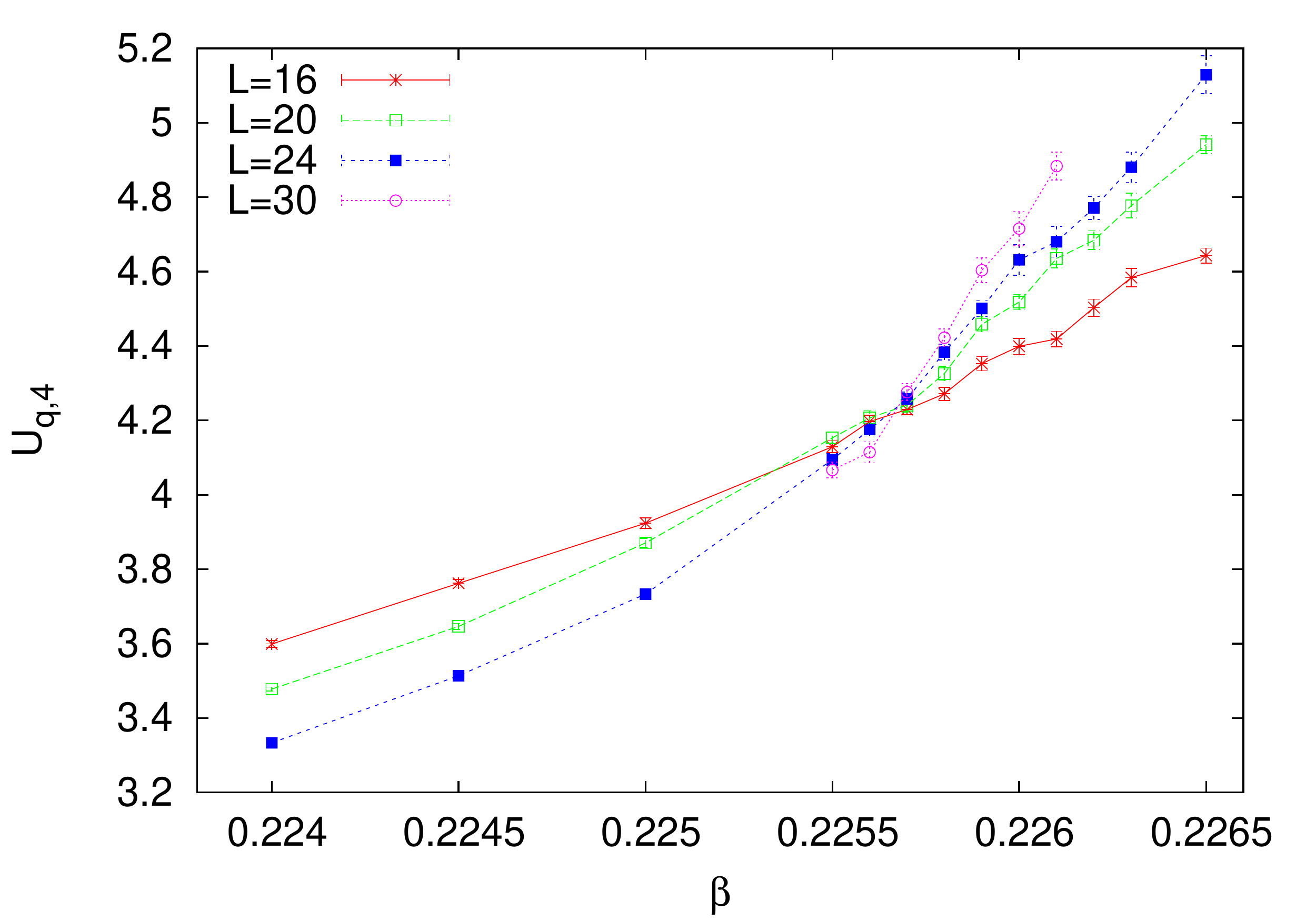}
\caption{Plot of $R_\xi = \xi_q/L$ (left) and of 
$U_{q,4}$ (right) versus $\beta$ for several values of $L$. 
All results are obtained for $\beta_G=\beta_{G}(L)$.  }
\label{fig:case-b-fit}
\end{figure*}

\begin{table}[tbph]
\begin{center}
\begin{tabular}{rcccccc }
\hline\hline
 & $n$ & $L_{\rm min}$ & $\beta_{c}$&$R^*$ &$\nu$ &$\chi^2$/DOF \\
			\hline
$R_\xi$   & 3 & 16 & 0.225980(12) & 0.1157(3) & 0.651(14) & 1.6 \\
          & 4 & 16 & 0.225977(12) & 0.1158(3) & 0.649(14) & 1.6 \\
          & 3 & 20 & 0.225951(16) & 0.1149(4) & 0.630(20) & 1.8 \\
          & 4 & 20 & 0.225953(17) & 0.1150(4) & 0.639(22) & 1.8 \\
$U_{q,4}$ & 3 & 16 & 0.225629(13) & 4.209(10) & 0.681(20) & 2.0 \\
          & 4 & 16 & 0.225629(13) & 4.210(10) & 0.670(21) & 1.9 \\
          & 3 & 20 & 0.225711(18) & 4.288(17) & 0.643(30) & 1.3 \\
          & 4 & 20 & 0.225708(19) & 4.289(18) & 0.655(33) & 1.2 \\
\hline\hline
\end{tabular}
\end{center}
\caption{Results [data at $\beta_G = \beta_{G}(L)$] of the fits to 
Eq.~(\ref{fits-R}): the order $n$ of the expansion is
given in the second column.
In each fit we only include the data satisfying $L\ge L_{\rm min}$.
In the last column we report the sum of the residuals divided by the 
number of degrees of freedom (DOF) of the fit.
}
\label{table:results-fits-bGL}
\end{table}

Let us now consider the behavior of the system at the 
finite-size pseudocritical transition $\beta_G(L)$. 
The data for $R_\xi = \xi_q/L$ and 
$U_{q,4}$ are shown in figure~\ref{fig:case-b-fit}. Note that in this case
we do not have data for $L=34$, we are considering a significantly
smaller interval of values of $\beta$, and the the number of samples is 
significantly smaller than for $\beta_G = \beta_{G,c}$. 
Therefore, we will use the results at $\beta_{G}(L)$ only 
as a consistency check of those obtained at the critical point, and 
in particular of the final estimates (\ref{final-est}).

As expected, data show an intersection 
for $\beta \simeq 0.226$, in agreement with the 
general theory. We analyze $R_\xi$ and $U_{q,4}$ as before, performing fits to 
Eq.~(\ref{fits-R}). Note that the 
constant $R^*$ should take here a value which is different  
from that it has for $\beta_G= \beta_{G,c}$, as here 
$R^* = f_R(X_{\rm max},0)$, see Eq.~(\ref{R-RG}). 
Results are reported in Table~\ref{table:results-fits-bGL}. 

The quality of the fits of $R_\xi$ is poor, with no improvement 
when data with $L=16$ are discarded. 
While the estimates of $\nu$ are consistent within errors with 
(\ref{final-est}),
the estimates of $\beta_c$
are significantly larger (by several error bars) than that obtained at
the critical point $\beta_{G,c}$. The results obtained from the 
analysis of $U_{q,4}$ are apparently better. For $L_{\rm min} = 20$,
the quality of the fits is reasonable and the estimates of 
$\nu$ and $\beta_c$ are both consistent with (\ref{final-est}).

We do not have enough data here to perform
an analysis that includes scaling corrections with an unconstrained exponent 
$\omega$, as performed at $\beta_{G,c}$. 
Stable results are only obtained if we fix $\omega$. 
We fix 
it to the central estimate (\ref{final-est}), $\omega = 1.0$, and we 
perform a combined analysis of the two observables to 
(\ref{fits-R2}). Using $n = 4$ and $m=2$ for both observables, we obtain 
$\beta_c = 0.225857(33)$ and $\nu = 0.66(10)$ with $\chi^2$/DOF$ \approx 1.3$. 
If we decrease $\omega$ by one error bar, the estimates change by much less than
the quoted errors. The estimate of $\beta_c$ is essentially 
consistent within error 
bars with Eq.~(\ref{final-est}). This analysis clearly confirms that 
the discrepancy observed for $\beta_c$ in the analysis of $R_\xi$ 
can be ascribed to the neglected scaling corrections.

\begin{figure}[!tbph]
\begin{tabular}{cc}
\includegraphics[width=0.5\textwidth,keepaspectratio]{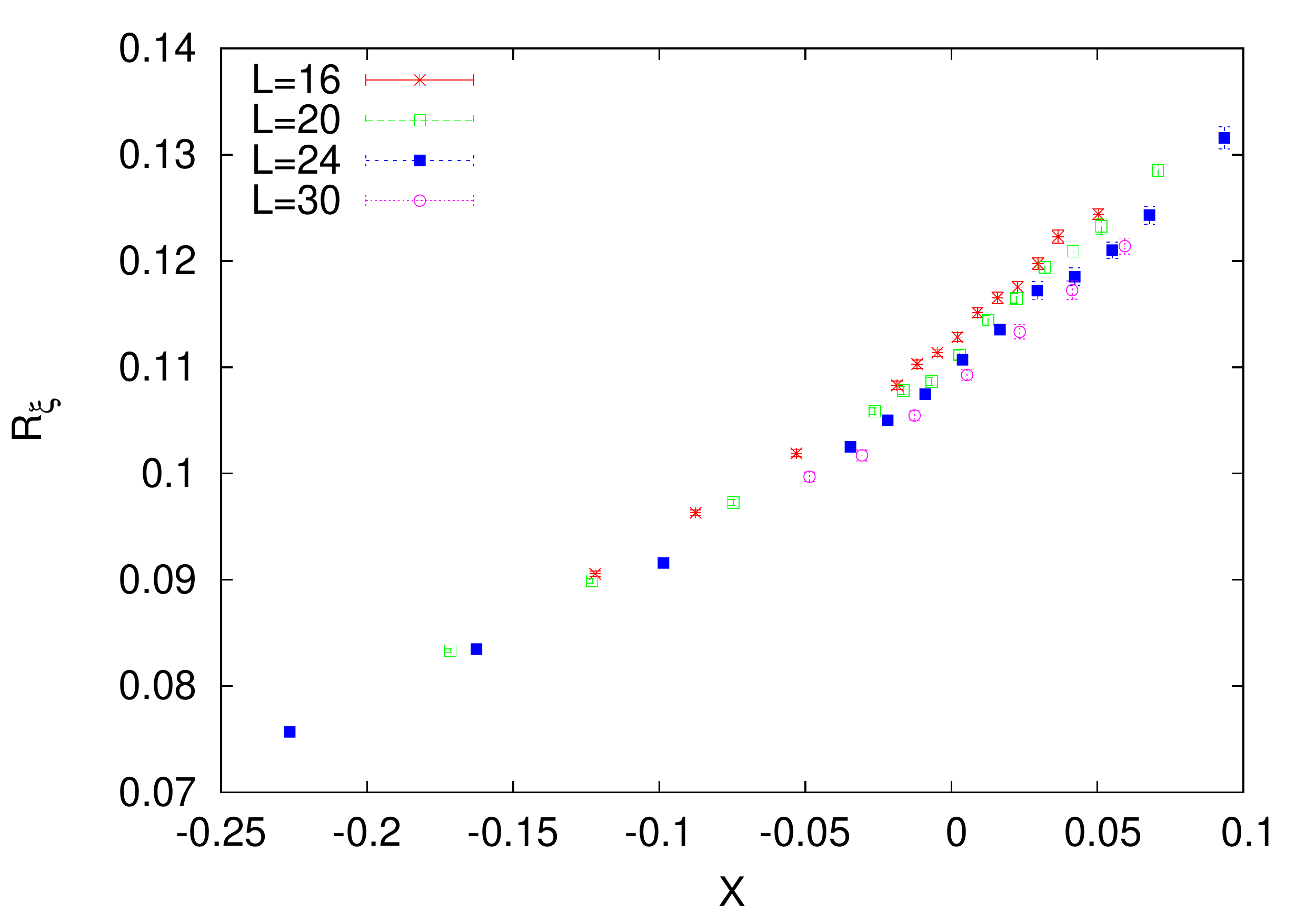}
&
\includegraphics[width=0.5\textwidth,keepaspectratio]{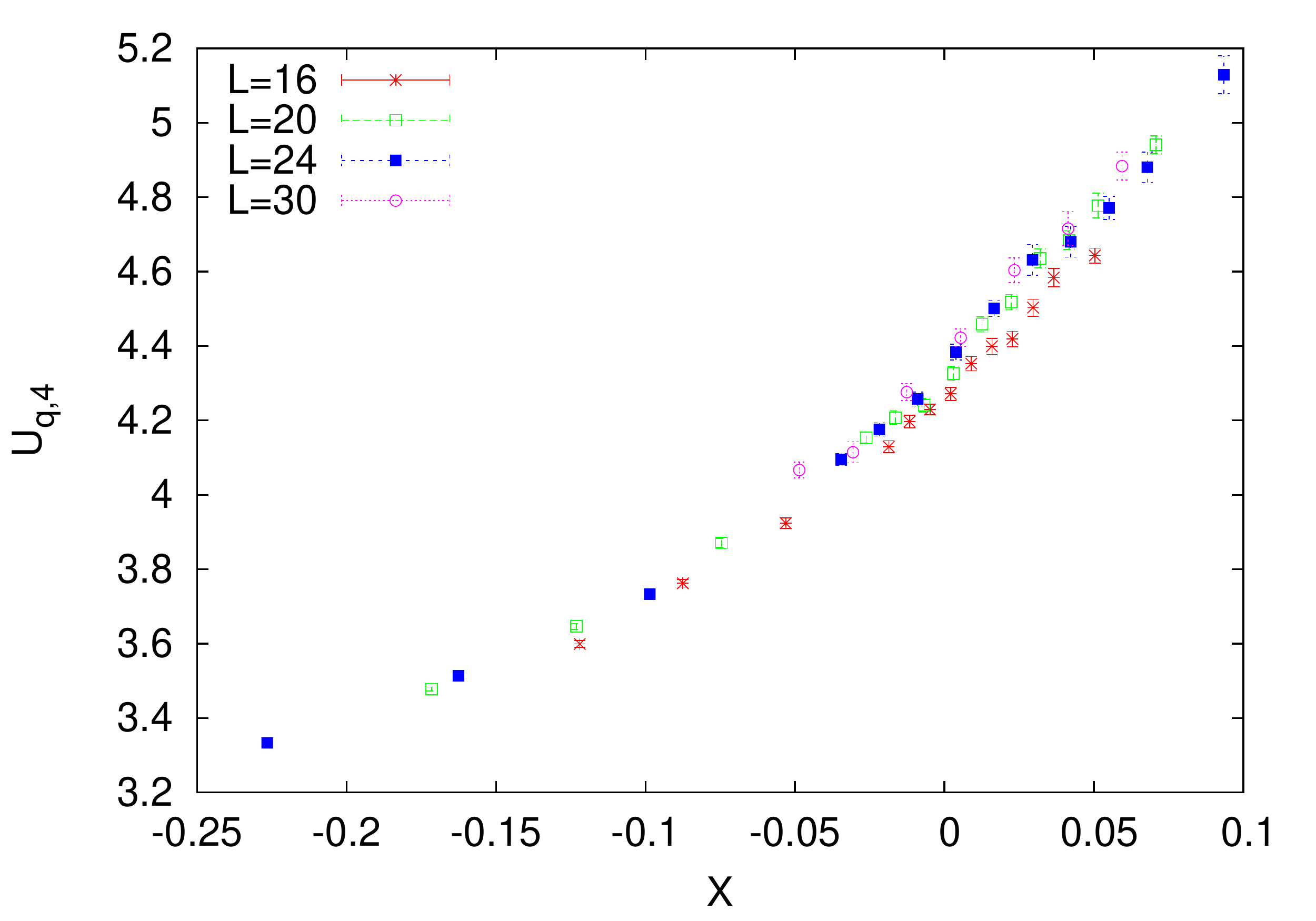}
\\
\end{tabular}
\caption{Results for $R_\xi = \xi_q/L$ (left) and $U_{q,4}$ (right) 
as a function of $X = (\beta - \beta_c) L^{1/\nu}$ for several values of $L$. 
All results for $\beta_G=\beta_{G}(L)$. We set 
$\beta_c = 0.22577$ and $\nu = 0.655$. 
}
\label{fig:case-b-collapse}
\end{figure}

We show the data as a 
function of $X = (\beta-\beta_c)L^{1/\nu}$ in 
figure~\ref{fig:case-b-collapse}, using the estimates 
(\ref{final-est}). Scaling corrections are clearly visible, but data are 
consistent with a universal scaling.

\subsection{Overlap susceptibility exponent}  \label{sec:etaq}

It is also interesting to analyze the behavior of the susceptibility 
$\chi_q$ to determine the exponent $\eta_q$. For $\beta_G = \beta_{G,c}$ 
we can expand the scaling function defined in~(\ref{chi-RG}) in 
powers of $X$, to obtain
\begin{equation}
\ln \chi_q = (2 - \eta_q) \ln L + \sum_{k=0}^n a_k X^k + b (\beta - \beta_c).
\label{fitchi}
\end{equation}
The last term, proportional to $(\beta - \beta_c)$ represents the first 
term in the expansion of the nonlinear scaling field $u_\chi$. 
If we fix $\beta_c$ and $\nu$ to the central values (\ref{final-est}), we obtain
$\eta_q = 1.055(6)$, 1.059(8), 1.038(15) for $L_{\rm min} = 16,20,24$,
respectively. If we change $\nu$ by one error bar, results are essentially
unchanged, while they vary by less than 0.04 
if we change $\beta_c$ by the same amount. 
Thus, we end up with the final estimate 
\begin{equation}
    \eta_q = 1.05(5).
\label{eta-final}
\end{equation}
We have also performed fits including scaling corrections.
For the susceptibility there are two sources of corrections. First, there 
are the nonanalytic corrections that decay as $L^{-\omega}$, as discussed for 
the RG invariant quantities. Second, there are corrections due to the 
background that scale as $L^{\eta_q-2}$. The exponents $\omega$ and 
$\eta_q - 2$ are
close and it is not possible to include them both in the 
fitting function. Therefore, we have performed two different fits:
\begin{equation}
\ln \chi_q = (2 - \eta_q) \ln L + \sum_{k=0}^n a_k X^k + b (\beta - \beta_c) + 
           L^{-\omega} \sum_{k=0}^m c_k X^k,
\label{fitchi-corr1}
\end{equation}
and 
\begin{equation}
\ln \chi_q = (2 - \eta_q) \ln L + \sum_{k=0}^n a_k X^k + b (\beta - \beta_c) + 
           L^{\eta_q - 2} \sum_{k=0}^m c_k (\beta - \beta_c)^k.
\label{fitchi-corr2}
\end{equation}
Using the central values for $\beta_c$, $\nu$, and $\omega$ [see
(\ref{final-est})], $L_{\rm min} = 16$ and 
$n = 6$, $m=2$, fits (\ref{fitchi-corr1}) 
and (\ref{fitchi-corr2}) both give $\eta_q = 1.04(5)$. Results are 
perfectly consistent with (\ref{eta-final}). 

The value of $\eta_q$ is close to the value it assumes in the ferromagnetic
Ising model. Indeed, the exponent $\eta_q$ for 
the overlap susceptibility is related to the exponent $\eta_m$ for the 
magnetic susceptibility by $\eta_q = 1 + 2 \eta_m$, so that for the 
Ising model we have
$\eta_q = 1.0725(2)$ [we use \cite{Hasenbusch-10} $\eta_m=0.03627(10)$]. 

As a consistency check, we have analyzed the data obtained setting 
$\beta_G = \beta_G(L)$. The quality of the fits to (\ref{fitchi}) 
is relatively poor,
$\chi^2$/DOF$ \approx 1.9$. We obtain $\eta_q = 1.226(3)$ and 
$\eta_q = 1.217(4)$ for $L_{\rm min} = 16,20$, respectively.
The results are significantly different from those obtained 
at $\beta_{G,c}$ at the level of the statistical errors. 
This clearly indicates that there are significant scaling corrections. 
We have therefore performed fits including scaling corrections.
Using the central values for $\beta_c$, $\nu$, and $\omega$ [see
(\ref{final-est})], fits (\ref{fitchi-corr1}) and
(\ref{fitchi-corr2}) both give $\eta_q = 1.13(3)$. In both 
cases $\chi^2$/DOF is approximately 1.4. The result is essentially 
consistent with (\ref{eta-final}) once we take into account the statistical
errors.

\subsection{Overlap distribution}

\begin{figure}[!tbph]
\begin{center}
\begin{tabular}{cc}
\includegraphics[width=0.5\textwidth,keepaspectratio]{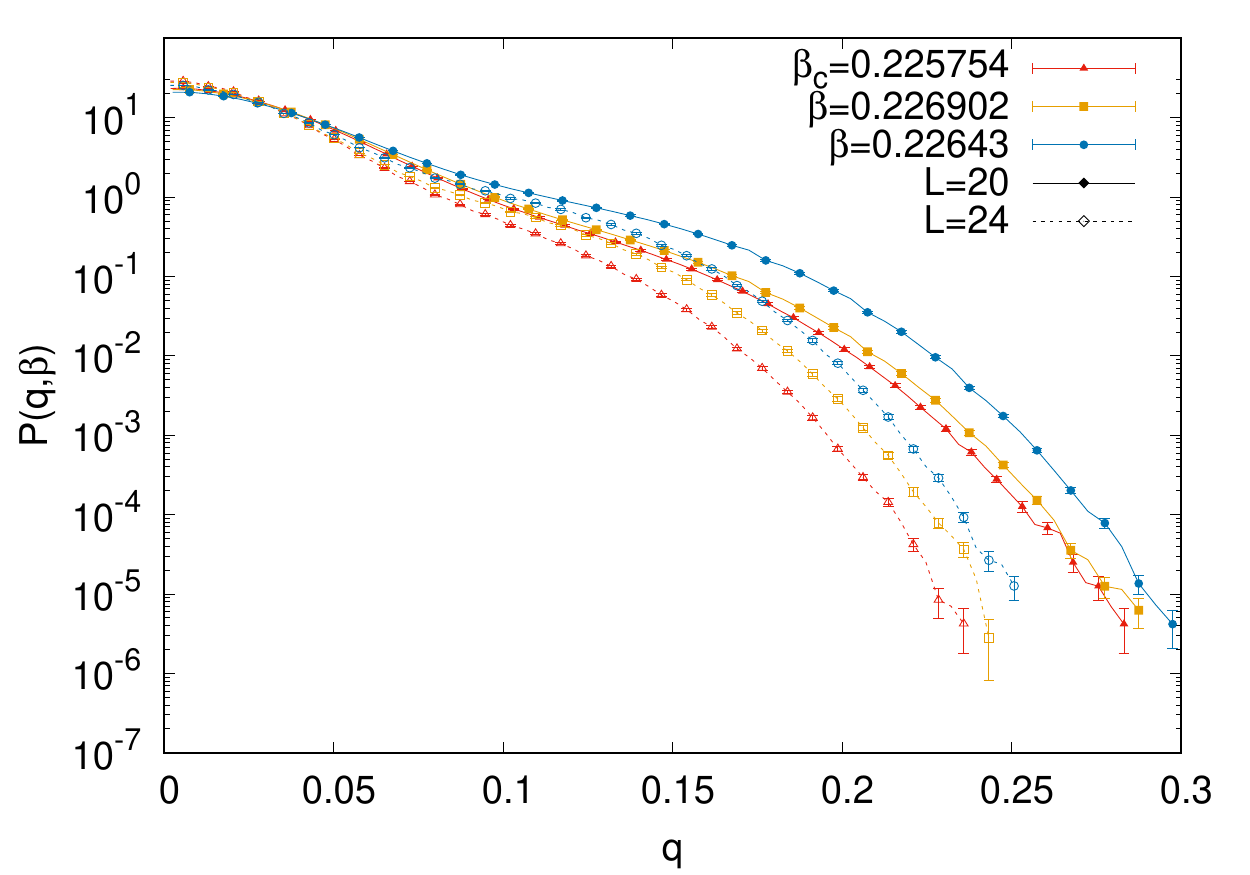}
& 
\includegraphics[width=0.5\textwidth,keepaspectratio]{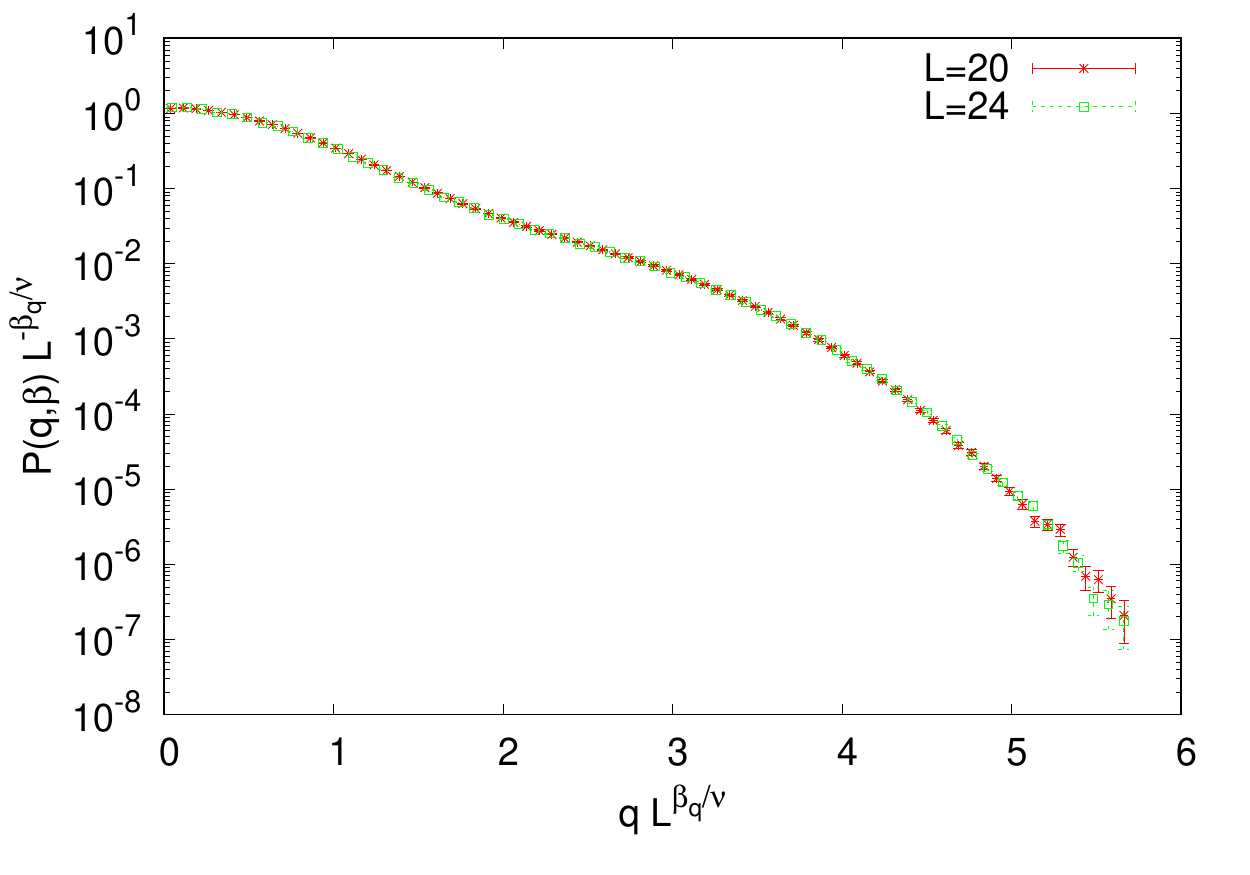}
\\
\end{tabular}
\end{center}
\caption{Left: Overlap distribution 
as a function of $q$ for three values of 
$\beta$: the smallest one corresponds to the transition point $\beta_c$.
Right: scaling plot of the overlap distribution as a function of 
$q L^{\beta_q/\nu}$ for $\beta = \beta_c$.
The distribution $P(q,\beta)$ is symmetric in $q$ 
and therefore we only report it for 
$q \ge 0$. In all cases $\beta_G = \beta_{G,c}$ and $L=20,24$.
}
\label{fig:Pq}
\end{figure}

To better characterize the  nature of the transition points we have 
determined the overlap distribution defined by 
\begin{equation}
    P(q,\beta) = [\langle \delta(Q/V - q)\rangle ],
\end{equation}
where $Q = \sum_{x} q_x$.
The results for $L=20$ and 24 and three slightly different values of 
$\beta$ at $\beta_G = \beta_{G,c}$
are reported in figure~\ref{fig:Pq}. 
The distribution is distinctly different from a Gaussian.
There is
a significant tail whose importance increases, as $\beta$ increases.
This tail is responsible for a critical value $U^*_{q,4}$ significantly 
larger than 3 (the Gaussian value). The function $P(q,\beta)$ should satisfy 
a general scaling relation. We define the overlap magnetization exponent
$\beta_q$, which is related to $\eta_q$ by the scaling relation 
\begin{equation}
{\beta_q\over \nu} = {1\over 2}(d-2 + \eta_q),
\end{equation}
where $d$ is the space dimension ($d=3$ in our case). Using 
(\ref{eta-final}) we find 
\begin{equation}
{\beta_q\over \nu} = 1.025(25).
\end{equation}
In terms of $\beta_q$ we have 
\begin{equation}
P(q,\beta) = L^{\beta_q/\nu} f_P(q L^{\beta_q/\nu}, X),
\end{equation}
where $f_P(x,X)$ is a universal (apart from a 
rescaling of its arguments) function, satisfying 
\begin{equation}
\int_{-\infty}^{\infty} f_P(x,X) dx = 1.
\end{equation}
In figure~\ref{fig:Pq} we report a scaling plot for $\beta = \beta_c$ 
(i.e., for $X=0$). Data for two different values of $L$ fall one on top
of the other, confirming the correctness of our estimates of $\eta_q$ and 
$\beta_q$. 

\begin{figure}[!tbph]
\begin{center}
\begin{tabular}{c}
\includegraphics[width=0.5
\textwidth,keepaspectratio]{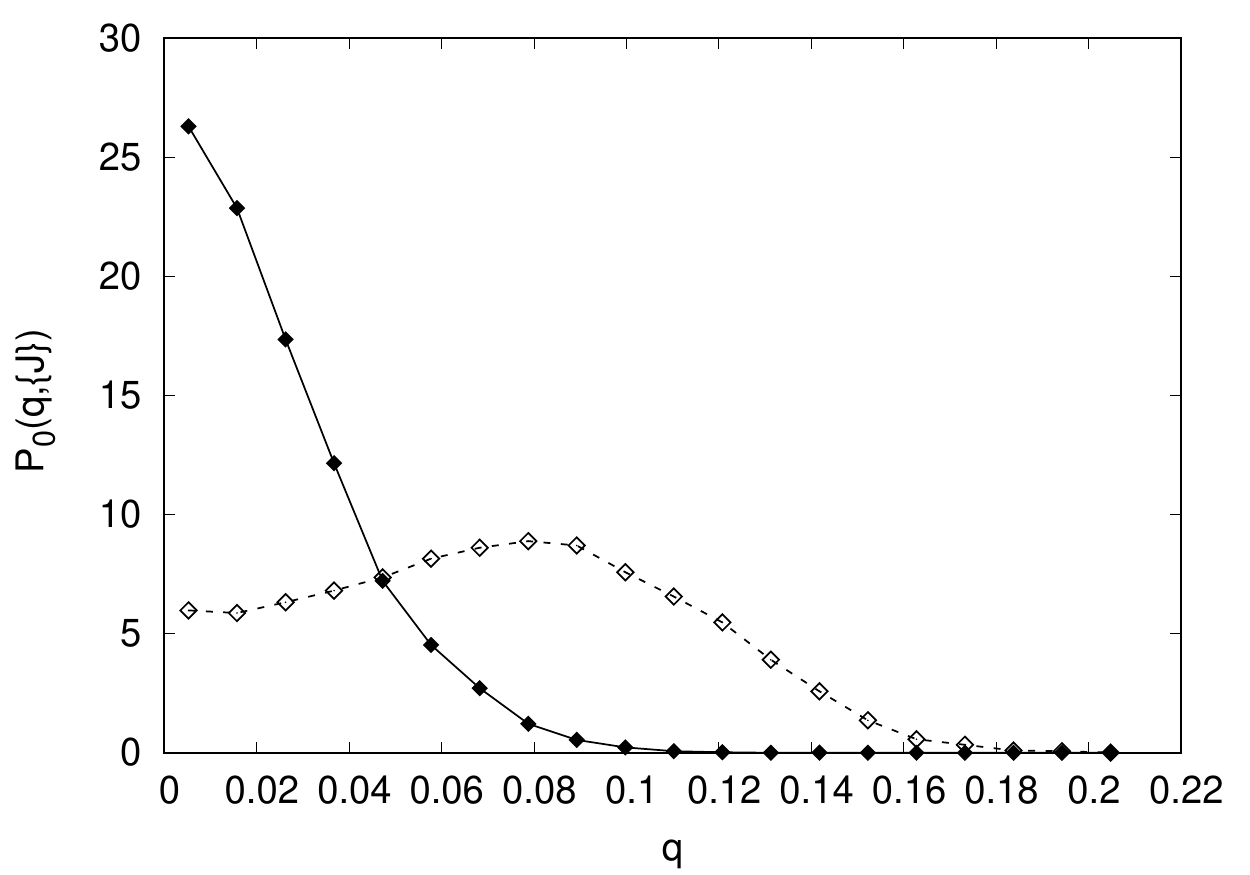}
\\
\end{tabular}
\end{center}
\caption{Overlap distribution $P_0(q,\{J\})$
as a function of $q$ ($\beta = \beta_c$, $\beta_G = \beta_{G,c}$) for two
different bond configurations.}
\label{fig:P0q}
\end{figure}

The overlap distribution has a peculiar shape that seems to indicate the 
presence of two different behaviors, depending on the disorder
distribution. To obtain a better understanding, consider the 
sample-dependent distribution
\begin{equation}
P_0 (q,\{J\}) = \langle \delta(Q/V - q) \rangle,
\end{equation}
where we only perform the thermal average. The results of figure~\ref{fig:Pq}
are consistent with the idea that, for most bond configurations, 
$P_0 (q,\{J\})$ is peaked around $q=0$. However, the shape of the tail seems 
to indicate the presence of rare configurations for which $P_0 (q,\{J\})$ 
is quite different. We have looked 
at the distributions $P_0 (q,\{J\}) $ for several bond distributions and we
have indeed identified two different typical shapes, see figure~\ref{fig:P0q}.
In most of the cases $P_0 (q,\{J\})$ is clearly peaked around $q=0$, but in 
a few cases the distribution is broader with a peak for $q\not=0$. 

\begin{figure}[!tbph]
\begin{center}
\begin{tabular}{cc}
\includegraphics[width=0.5\textwidth,keepaspectratio]{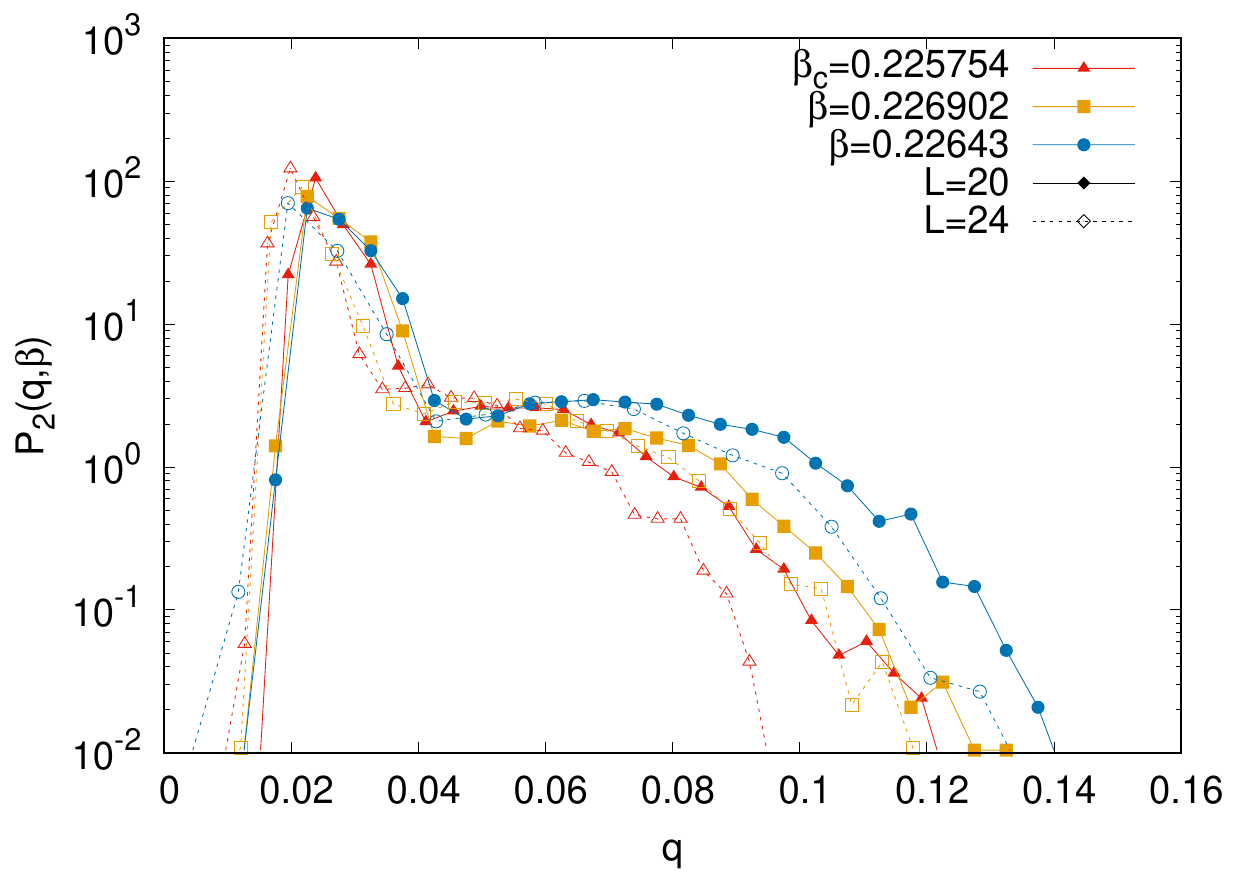}
& 
\includegraphics[width=0.5\textwidth,keepaspectratio]{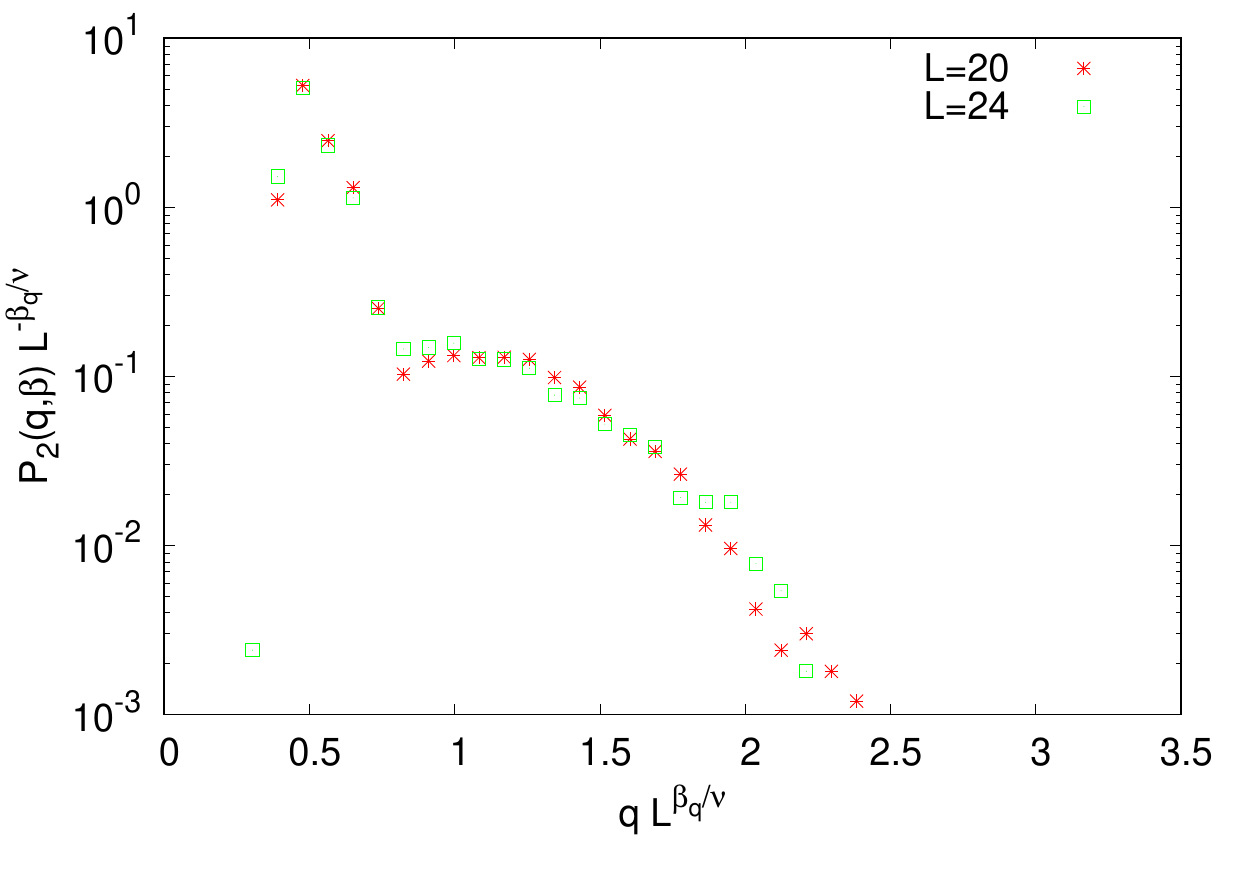}
\\
\end{tabular}
\end{center}
\caption{Left: distribution of $\langle |Q|\rangle/V$
for three values of 
$\beta$: the smallest one corresponds to the transition point $\beta_c$.
Right: scaling plot as a function of 
$q L^{\beta_q/\nu}$ for $\beta = \beta_c$.
In all cases $\beta_G = \beta_{G,c}$ and $L=20,24$.
}
\label{fig:Pq2}
\end{figure}

To better 
single out the presence of two different types of configurations we compute 
\begin{equation}
P_2 (q,\beta) = [\langle \delta(\langle |Q|\rangle /V - q) \rangle].
\end{equation}
As we consider $\langle |Q|\rangle /V$, we expect a sharp peak corresponding to
bond configurations such that $P_0 (q,\{J\})$ is peaked in $q=0$ and 
a wide tail that gets contributions from configurations with a peak for
$q\not=0$.
The results, shown in Fig.~\ref{fig:Pq2}, confirm 
the general analysis. In figure~\ref{fig:Pq2}, we also report a scaling plot,
that confirms the asymptotic nature of the results.

\section{Low-frustration regime} \label{sec5}

In the low-temperature phase of the gauge model, that is for 
$\beta_G>\beta_{G,c}$, frustration is suppressed. 
In order to study the intermediate region between $\beta_{G,c}$ and 
the zero-frustration limit $\beta_G=\infty$, we perform simulations for 
$\beta_G=0.9$, considering $16\le L \le 30$. For this value of $\beta_G$ 
sample-to-sample fluctuations are small and therefore we obtain 
significantly more precise results than at $\beta_G = \beta_{G,c}$.
For $\beta_G=0.9$, the fraction of frustrated plaquettes
is approximately $0.4\%$, with a weak $L$ dependence. 

\begin{figure}[!tbph]
\begin{tabular}{cc}
\includegraphics[width=0.5\textwidth,keepaspectratio]{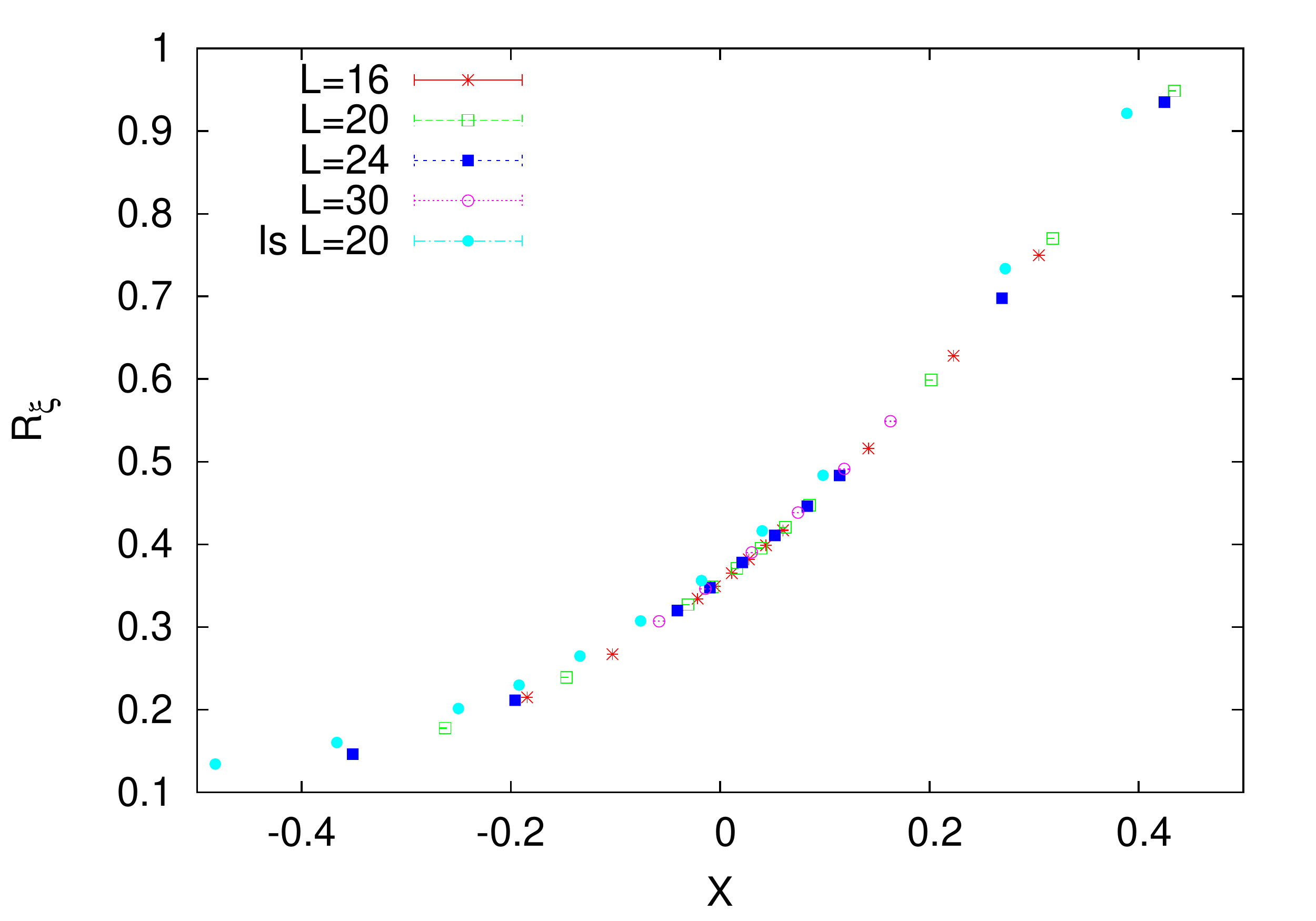}
&
\includegraphics[width=0.5\textwidth,keepaspectratio]{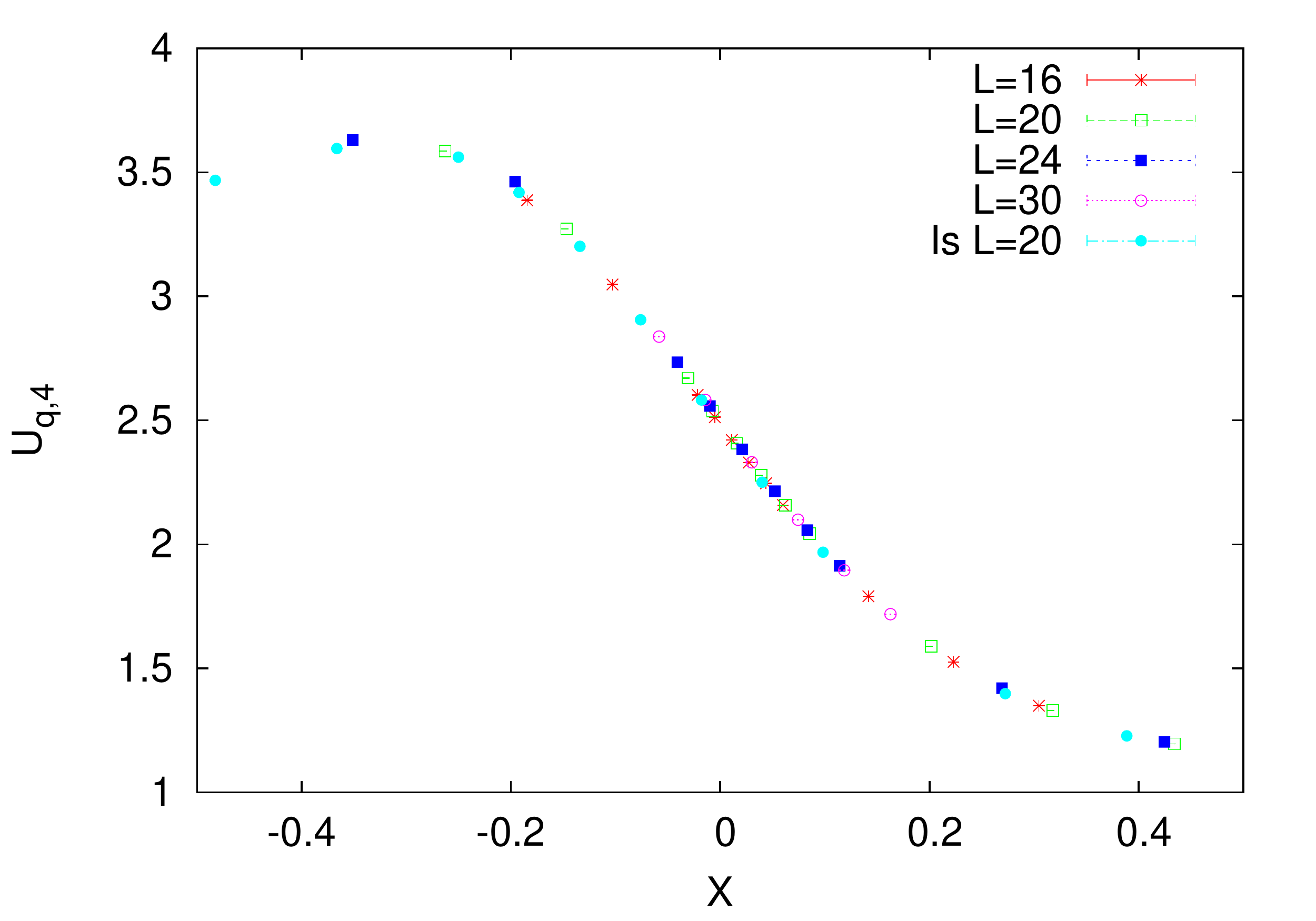}
\\
\end{tabular}
\caption{Scaling plot for $R_\xi = \xi_q/L$ (left) and 
$U_{q,4}$ (right) versus $(\beta - \beta_c) L^{1/\nu_I}$. We report data for 
the disordered model at $\beta_G=0.9$ and for the pure 
Ising model. We use 
$\nu_I=0.63002$ for both models, $\beta_c = 0.222264$ for the disordered model, 
$\beta_c = \beta_{I,c} = 0.2216547$ for the pure Ising model.}
\label{fig:q-09-collapse}
\end{figure}

The results for $\xi_q/L$ and $U_4^q$ have been analyzed as in the 
previous sections. The $\chi^2$ of the fits without scaling corrections is
large, so that we have been forced to include them in the analysis.
Combined fits of the two observables to (\ref{fits-R2}) (we use 
$n=8$ and $m=2$ for both observables) give
$\beta_c=0.222234(2)$, $\nu = 0.628(2)$, and $\omega=0.99(15)$. 
The estimates of 
$\nu$ and $\omega$ are very close to the pure Ising values \cite{Hasenbusch-10}
$\nu_I = 0.63002(10)$ and $\omega_I = 0.832(6)$. 
These results make us conjecture that the critical 
behavior of the model for $\beta_G = 0.9$ is in the same universality class as 
that of the pure Ising model. If this conjecture is correct, we can obtain a
more precise estimate of the critical point, reanalyzing the data fixing 
$\nu$ and $\omega$ to the Ising values of Ref.~\cite{Hasenbusch-10}. 
We obtain $\beta_c = 0.222264(4)$ including all data, and 
$\beta_c = 0.222262(9)$ if we only consider data with $L\ge 20$.
To provide stronger evidence for the 
conjecture, we compare in figure~\ref{fig:q-09-collapse} 
the scaling functions in the two models: they should 
agree, apart from a rescaling of the abscissa. For the Binder cumulant
$U_{4,q}$, data fall on top of each other with good accuracy, without
the need of any rescaling. On the other hand, the data for $R_\xi$ in the two
models show a tiny difference that might be due to scaling corrections, that 
are quite sizable for the range of values we consider, as we discussed. 

The results shown here apparently indicate that 
a small amount of (correlated) frustration does not change 
the universality class: The system at $\beta_G = 0.9$ behaves 
as the pure Ising model. However, the reader should be aware that lattices 
are quite small and, therefore, conclusive evidence can only be 
obtained by considering significantly larger lattices. This 
check should be feasible as 
the ferromagnetic nature of the transition shoud make 
cluster algorithms \cite{SW-87} quite efficient.

\section{Conclusions} \label{sec6}

In this paper we have considered a disordered Ising model in which the 
amount of frustration can be continuously varied in a gauge-invariant way
by tuning a single parameter $\beta_G$, 
that plays the role of a chemical potential
for the frustrated plaquettes. By changing $\beta_G$ we can interpolate 
between the usual spin-glass Edwards-Anderson model ($\beta_G = 0$) and the 
pure Ising model. However, the resulting phase diagram is quite different
from that observed in the usual $\pm J$ model with bond distribution 
(\ref{prob-EA}). In that model, for $p\le 1/2$, there are only two  
distinct phases, a ferromagnetic one in the same universality class as 
that of the randomly dilute Ising (RDI) model  
and a spin-glass phase separated by a 
multicritical point. The phase diagram of the model we consider here is 
apparently
richer. For $\beta_G = \beta_{G,c}$ we identify a new universality class
characterized by the fact that the distribution of frustration is 
long-range correlated. We compute the correlation-length exponent $\nu$ and 
the overlap-susceptibility exponent $\eta_q$, obtaining
\begin{equation}
\nu = 0.655(15), \qquad \eta_q = 1.05(5)
\end{equation}
Given the very small amount of frustration (the fraction of frustrated
plaquettes is 2.6\% for $\beta_G = \beta_{G,c}$), 
we expect the transition to 
be ferromagnetic in an appropriate set of variables. Note that, because of 
gauge invariance, one cannot consider correlations of the spins. For instance,
it is trivial to show that $[\langle \sigma_x\sigma_y\rangle] = \delta_{xy}$. 
One possibility consists in considering gauge-invariant observables of the 
form 
\begin{equation}
f_{xy}(\mathcal{C}) = \langle \sigma_x \prod_{l \in \mathcal{C}}J_{l}\,
\sigma_y \rangle,
\label{def-f-C}
\end{equation}
where $\mathcal{C}$ is a lattice path that connects sites $x$ and $y$.
When $\beta_G = \infty$, we recover the usual definition of the
magnetic correlation function in an appropriate gauge. Indeed, in this limit
$\Pi_P = 1$ on all plaquettes and therefore by means of a gauge transformation
we can set $J_{xy} = 1$ on all links. In this gauge the Hamiltonian of the
system is that of the pure Ising model and the correlation function becomes
$\langle\sigma_x\sigma_y\rangle$. Observables defined as in 
(\ref{def-f-C}) were already considered in
\cite{APV-08}, where it was shown that they had the expected scaling 
behavior provided the path $\mathcal{C}$ was appropriately chosen.

For $\beta_G > \beta_{G,c}$ our preliminary data 
are consistent with pure Ising behavior, but significant larger systems 
are needed to draw a definite conclusion. The behavior for $\beta_G <
\beta_{G,c}$ has not been investigated yet. One possibility is that 
in the whole interval $0 \le \beta_G < \beta_{G,c}$ there is only a spin
glass phase. However, at $\beta_{G,c}$ frustration is quite small: 
the percentage of frustrated plaquettes is only 2.6\%, which is significantly
smaller than the fraction, 46.2\%, at $p^*$ for the EA model with 
disorder distribution (\ref{prob-EA}). Thus, it is possible 
that in this interval the behavior is analogous to that observed in 
the $\pm J$ model with uncorrelated disorder: a spin-glass phase 
for $\beta_G < \beta_{G,MC}$ and a ferromagnetic RDI phase for 
$\beta_{G,MC} < \beta_G < \beta_{G,c}$. Finally, it would be interesting 
to investigate the behavior for $\beta_G < 0$. For $\beta_G = -\infty$ 
we would obtain the fully frustrated Ising model. For the cubic lattice, 
all numerical simulations \cite{CFH-82,DLN-85,Grest-85,BHT-99,MKR-05} 
are consistent with the presence of
a second-order high-temperature transition (however, so far, 
the transition has not been explained within the Landau-Ginzburg-Wilson
approach, as no appropriate fixed point has been found \cite{BMNB-84}) 
and of a first-order low-temperature transition. 
It is tempting to assume that the two transitions survive if we take 
$|\beta_G|$ large but finite. In this case, the phase diagram would be 
quite complex as the two transition lines should necessarily 
merge as $|\beta_G|$ decreases.

\medskip

The simulations presented here have been performed on the Theory cluster 
at INFN, Sezione di Roma 1. 

\appendix
\section{Computational details} 

\begin{table}
\caption{Number of samples $N_s$ and number of iterations per sample $N_{m}$ 
for the runs at $\beta_G = \beta_{G,c}$.}
\label{data1}
\begin{center}
\begin{tabular}{cccc cccc}
\hline\hline
$L$ & $\beta$ & $N_s/10^5$ & $N_m/10^3$ & 
$L$ & $\beta$ & $N_s/10^5$ & $N_m/10^3$ \\
\hline
\footnotesize
16 & 0.2200 & 3.2  & 10 & 24 &   0.2257 & 19.2   & 5.5 \\
 & 0.2220 & 3.2    & 10        & & 0.2258& 19.2  & 5.5  \\
 & 0.2240 & 25.6   & 10       & & 0.2259 & 19.2   & 5.5  \\
 & 0.2245 & 12.8   & 10       & & 0.2260 & 11.5   & 5.5  \\
 & 0.2250 & 12.8   & 10       & & 0.2262 & 12.8   & 5.5  \\
 & 0.2255 & 12.8   & 10       & & 0.2264 & 12.8   & 5.5  \\
 & 0.2260 & 25.6   & 10       & & 0.2265 & 10.2   & 5.5  \\
 & 0.2265 & 12.8   & 10       & & 0.2267 & 12.8   & 5.5  \\
 & 0.2270 & 2.56   & 10       & & 0.2270 & 3.2   & 10   \\
 & 0.2280 & 3.2    & 10       & & 0.2280 & 3.2   & 10  \\
 & 0.2300 & 3.2    & 10 & 30    & 0.2230 & 3.2   & 10  \\
20 & 0.2200 & 0.64 & 10       & & 0.2240 & 3.8   & 30  \\
 & 0.2220 & 0.64   & 10       & & 0.2245 & 3.8   & 30  \\
 & 0.2240 & 11.8   & 10       & & 0.2250 & 3.8   & 30  \\
 & 0.2245 & 12.8   & 10       & & 0.2255 & 19.2   & 5.5  \\
 & 0.2250 & 12.8   & 10       & & 0.2256 & 19.2   & 5.5  \\
 & 0.2255 & 19.2   & 10       & & 0.2257 & 19.2   & 5.5  \\
 & 0.2256 & 19.2   & 10       & & 0.2258 & 19.2   & 5.5  \\
 & 0.2257 & 19.2   & 5        & & 0.2259 & 19.2   & 5.5  \\
 & 0.2258 & 19.2   & 5        & & 0.2260 & 6.4   & 5.5  \\
 & 0.2259 & 19.2   & 5        & & 0.2262 & 6.4   & 5.5  \\ 
 & 0.2260 & 19.2   & 5        & & 0.2264 & 6.4   & 5.5  \\
 & 0.2262 & 12.8   & 5        & & 0.2265 & 6.4   & 5.5  \\
 & 0.2264 & 19.2   & 5        & & 0.2267 & 6.4   & 5.5  \\
 & 0.2265 & 19.2   & 5        & & 0.2270 & 6.4   & 5.5  \\
 & 0.2267 & 19.2   & 5  & 34    & 0.2255 & 12.8   & 5.5  \\
 & 0.2270 & 16     & 5        & & 0.2256 & 12.8   & 5.5  \\
 & 0.2280 & 3.2    & 10       & & 0.2257 & 12.8   & 5.5  \\
 & 0.2300 & 3.2    & 10       & & 0.2258 & 12.8   & 5.5  \\
24 & 0.2200 & 0.64 & 10       & & 0.2259 & 12.8   & 5.5  \\
 & 0.2220 & 0.64   & 10       & & 0.2260 & 6.4   & 5.5  \\
 & 0.2240 & 5.12   & 10       & & 0.2262 & 6.4   & 5.5  \\
 & 0.2245 & 5.12   & 10       & & 0.2264 & 6.4   & 5.5  \\
 & 0.2250 & 5.12   & 10       & & 0.2265 & 6.4   & 5.5  \\
 & 0.2255 & 19.2   & 5.5      & & 0.2267 & 6.4   & 5.5  \\
 & 0.2256 & 19.2   & 5.5      & & 0.2270 & 6.4   & 5.5  \\
\hline\hline
\end{tabular}
\end{center}
\end{table}

The Ising model (\ref{HamJ}) has been simulated using a standard Metropolis 
multispin code that deals with 64 systems at the same time, while the bond 
configurations have been generated using a standard Metropolis algorithm.
In practice, we first perform a simulation of the gauge model.
Every $N_{\rm it}$ iterations (one iteration consists in proposing an update 
of every link variable) a bond configuration is saved. Once 64
bond configurations have been generated, they are packed in a single 64-bit word
and a simulation of the spin system is performed. Two replicas are simulated 
together, in order to compute functions of the overlap.
The parameters of the different runs are reported in Tables \ref{data1} and 
\ref{data2}.

The parameter $N_{\rm it}$ has been chosen so that successive bond 
configurations are approximately independent. For this purpose, 
we have computed the autocorrelation time \cite{Sokal-97}
$\tau$ for the average energy and for the average of the so-called 
Polyakov loop $P$. The Polyakov loop is defined as the product of the bond 
variables associated with a topologically nontrivial line that is 
parallel to a coordinate direction and connects the opposite boundaries of the
lattice. The Polyakov loop is usually considered as the order parameter of
the transition in the gauge model. The dynamics of the Polyakov loop is 
slower than that of the energy and, for $\beta_G = \beta_{G,c}$, 
its correlation times $\tau$ is approximately $26,57,115,171$ 
for $L=10,14,18,23$ (here $\tau$ is the 
integrated autocorrelation time as defined in 
\cite{Sokal-97}). The results are well fitted by $0.12 L^{2.33}$.
Note that this is simply a phenomenological fit, which should be correct
in the interval of values of $L$ of interest. Indeed, for large $L$ 
we expect $\tau$ to increase with an exponent very close to 2
\cite{Grassberger-95}. 
In the calculations at $\beta_{G,c}$ we have taken $N_{\rm it} = \tau$, 
while in the runs at $\beta_G(L)$ we have taken $N_{\rm it} \approx \tau/2$.
Since we obtain the bond configurations in a dynamic Monte Carlo simulation, 
results
corresponding to different samples are not uncorrelated. To obtain 
correct estimates of the statistical errors, they have been determined using 
a blocking method. Errors have been obtained from the variance of averages 
over a large (typically of the order of 6400) number of samples.
In disordered systems, thermalization is crucial. We discarded in all cases 
640 iterations. With this choice there is no evidence of initialization bias. 
This has been verified in detail in the runs with $\beta_G = \beta_G(L)$,
which are significantly longer than needed.

\begin{table}
\caption{Number of samples $N_s$ and number of iterations per sample $N_{m}$ 
for the runs at $\beta_G = \beta_{G}(L)$.}
\label{data2}
\begin{center}
\begin{tabular}{cccc cccc}
\hline\hline
$L$ & $\beta$ & $N_s/10^5$ & $N_m/10^3$ & 
$L$ & $\beta$ & $N_s/10^5$ & $N_m/10^3$ \\
\hline
\footnotesize
16 & 0.2240 & 6.4  & 96 & 20 & 0.2262 & 4.48    & 140  \\
 & 0.2245 & 6.4    & 96 &    & 0.2263 & 4.48    & 140  \\
 & 0.2250 & 6.4    & 96 &    & 0.2265 & 6.4     & 140  \\
 & 0.2255 & 6.4    & 96 & 24 & 0.2240 & 2.24  & 180 \\
 & 0.2256 & 6.4    & 96 &    & 0.2245 & 2.62    & 180  \\
 & 0.2257 & 6.4    & 96 &    & 0.2250 & 2.82    & 180  \\
 & 0.2258 & 6.4    & 96 &    & 0.2255 & 6.4    & 180  \\
 & 0.2259 & 6.4    & 96 &    & 0.2256 & 6.4    & 180  \\
 & 0.2260 & 6.4    & 96 &     & 0.2257 & 6.4    & 180  \\
 & 0.2261 & 5.12    & 96 &   & 0.2258 & 6.4    & 180  \\
 & 0.2262 & 5.12    & 96 &   & 0.2259 & 6.4    & 180  \\
 & 0.2263 & 5.12    & 96 &   & 0.2260 & 2.62    & 180  \\
 & 0.2265 & 6.4    & 96 &    & 0.2261 & 2.56    & 180  \\
20 & 0.2240 & 6.4  & 140 &   & 0.2262 & 2.56    & 180  \\
 & 0.2245 & 6.4    & 140 &   & 0.2263 & 2.56    & 180  \\
 & 0.2250 & 6.4    & 140 &   & 0.2265 & 1.92    & 180  \\
 & 0.2255 & 6.4    & 140 &30 & 0.2255 & 3.2     & 30 \\
 & 0.2256 & 6.4    & 140 &   & 0.2256 & 3.2     & 30  \\
 & 0.2257 & 6.4    & 140 &   & 0.2257 & 4.28    & 75  \\
 & 0.2258 & 6.4    & 140 &   & 0.2258 & 4.28    & 75  \\
 & 0.2259 & 6.4    & 140 &   & 0.2259 & 3.2    & 30  \\
 & 0.2260 & 6.4    & 140 &   & 0.2260 & 3.2    & 30  \\
 & 0.2261 & 4.48    & 140 &  & 0.2261 & 3.2    & 30  \\
\hline\hline
\end{tabular}
\end{center}
\end{table}


\section*{References}

\begin{thebibliography}{99}

\bibitem{EA-75}
Edwards S F and Anderson P W 1975
Theory of spin glasses,
{\em J. Phys. F: Met. Phys.} {\bf 5} 965

\bibitem{IATKST-86}
Ito A, Aruga H, Torikai E, Kikuki M, Syono Y and Takei H 1986
Time-Dependent Phenomena in a Short-Range Ising Spin-Glass
Fe$_{0.5}$Mn$_{0.5}$ TiO$_3$, 
{\em Phys. Rev. Lett.} {\bf 57} 483

\bibitem{GSNLAI-91}
Gunnarsson K, Svedlindh P, Nordblad P, Lundgren L, Aruga H and Ito A 1991
Static scaling in a short-range Ising spin glass,
{\em Phys. Rev.} B {\bf 43} 8199

\bibitem{NN-07}
Nair S and Nigam A K 2007
Critical exponents and the correlation length in the manganite spin glass 
Eu$_{0.5}$Ba$_{0.5}$MnO$_3$,
{\em Phys. Rev.} B {\bf 75} 214415

\bibitem{Aharony-76}
Aharony A 1976
in {\em Phase Transitions and Critical Phenomena} vol. 6, ed C Domb and 
M S Green (New York: Academic) p 357 

\bibitem{PV-02-review}
Pelissetto A and Vicari E 2001
Critical Phenomena and Renormalization-Group Theory
{\em Phys. Rep.} {\bf 368} 549 


\bibitem{HPPV-07-pmj}
Hasenbusch M, Parisen Toldin F, Pelissetto A and Vicari E 2007
Critical Behavior of the Three-Dimensional
$\pm J$ Ising Model at the Paramagnetic-Ferromagnetic Transition Line,
{\em Phys. Rev.} B {\bf 76} 094402

\bibitem{BFMMPR-98}
Ballesteros H G, Fern\'andez L A, Mart{\'\i}n-Mayor V, 
Mun\~oz-Sudupe A, Parisi G and Ruiz-Lorenzo J J 1998 
Critical exponents of the three-dimensional diluted Ising model,
{\em Phys. Rev.} B {\bf 58} 2740

\bibitem{PV-02}
Pelissetto A and Vicari E 2002
Randomly dilute spin models: A six-loop field-theoretic study
{\em Phys. Rev.} B {\bf 62} 6393

\bibitem{HPPV-07}
Hasenbusch M, Parisen Toldin F, Pelissetto A and Vicari E 2007
Universality Class of 3D Site-Diluted and Bond-Diluted
Ising Systems,
{\em J. Stat. Mech.: Theory Expt.} P02016

\bibitem{KKY-06}
Katzgraber H, K\"orner M and Young A P 2006
Universality in three-dimensional Ising spin glasses: A Monte Carlo study,
{\em Phys. Rev.} B {\bf 73} 224432

\bibitem{HPV-08}
Hasenbusch M, Pelissetto A and Vicari E 2008
Critical behavior of three-dimensional Ising spin glass models,
{\em Phys. Rev.} B {\bf 78} 214205

\bibitem{Janus-13}
Baity-Jesi M {\em et al.} (Janus Collaboration) 2013
Critical parameters of the three-dimensional Ising spin glass
{\em Phys. Rev.} B {\bf 88} 224416

\bibitem{LPP-16}
Lulli M, Parisi G and Pelissetto A 2016
Out-of-equilibrium finite-size method for critical behavior analyses
{\em Phys. Rev.} E {\bf 93} 032126

\bibitem{Nishimori-81}
Nishimori H 1981
Internal energy, specific heat and correlation function of the bond random
Ising model,
{\em Prog. Theor. Phys.} {\bf 66} 1169

\bibitem{Nishimori-86}
Nishimori H 1986 
Geometry-induced phase transition in the $\pm J$ Ising model,
{\em J. Phys. Soc. Japan} {\bf 55} 3305

\bibitem{Nishimori-book}
Nishimori H 2001 {\em Statistical Physics of Spin Glasses and 
Information Processing: An Introduction}\/
(Oxford University Press, Oxford)

\bibitem{HPPV-07-multicr}
Hasenbusch M, Parisen Toldin F, Pelissetto A and Vicari E 2007
Magnetic-glassy multicritical behavior of the three-dimensional $\pm J$
Ising model,
{\em Phys. Rev.} B {\bf 76} 184202

\bibitem{BDI-75}
Balian R, Drouffe J M and Itzykson C 1975
Gauge fields on a lattice. II. Gauge invariant Ising model
{\em Phys. Rev.} D {\bf 11} 8

\bibitem{Kogut-79}
Kogut J B 1979
An introduction to lattice gauge theory and spin systems
{\em Rev. Mod. Phys.} {\bf 51} 4

\bibitem{PTPV-06}
Parisen Toldin F, Pelissetto A and Vicari E 2006
Critical behavior of the random-anisotropy model in the strong-anisotropy 
limit 
{\em J. Stat. Mech: Th. Expt.} P06002

\bibitem{Savit-80}
Savit R 1980
Duality in field theory and statistical systems,
{\em Rev. Mod. Phys.} {\bf 52} 453

\bibitem{Hasenbusch-10}
Hasenbusch M  2010
Finite size scaling study of lattice models in the three-dimensional Ising
universality class
{\em Phys. Rev.} B {\bf 82} 174433 

\bibitem{FXL-18}
Ferrenberg A M, Xu J and Landau D P 2018
Pushing the limits of Monte Carlo simulations for the three-dimensional Ising
model
{\em Phys. Rev.} E {\bf 97} 043301

\bibitem{KPSDV-16}
Kos F, Poland D, Simmons-Duffin D and Vichi A 2016
Precision islands in the Ising and $O(N)$ models
{\em J. High Ener. Phys.} P08036

\bibitem{SW-87}
Swendsen R H and Wang J S 1987
Nouniversal critical dynamics in Monte Carlo simulations
{\em Phys. Rev. Lett.} {\bf 58} 86

\bibitem{APV-08} 
Alba V, Pelissetto A, and Vicari E 2008
The Uniformly Frustrated Two-Dimensional $XY$ Model in the
Limit of Weak Frustration,
{\em J. Phys.} A: {\em Math. Theor.} {\bf 41} 175001


\bibitem{CFH-82}
Chui S T, Forgacs G and Hatch D M 1982
Ground states and the nature of a phase transition in a simple cubic
fully frustrated Ising model
{\em Phys. Rev.} B {\bf 25} 6952

\bibitem{DLN-85}
Diep H T, Lallemand P and Nagai O 1985
Critical properties of a simple cubic fully frustrated Ising lattice by Monte
Carlo method
{\em J. Phys. C: Solid State Phys.} {\bf 18} 1067

\bibitem{Grest-85}
G S Grest 1985 
Fully and partially frustrated simple cubic Ising models: a Monte Carlo study
{\em J. Phys. C: Solid State Phys.} {\bf 18} 6239

\bibitem{BHT-99}
Bernardi L W, Hukushima K and Takaya H 1999
Fully frustrated Ising system on a 3D simple cubic lattice: revisited
{\em J. Phys. A: Math. Gen.} {\bf 32} 1787

\bibitem{MKR-05}
Murtazaev A K, Kamilov I K, and Ramazanov M K 2005
Critical Properties of the Three-Dimensional Frustrated
Ising Model on a Cubic Lattice
{\em Phys. Solid State} {\bf 47} 1163

\bibitem{BMNB-84}
Blankschtein D, Ma M and Nihat Berker A 1984
Fully and partially frustrated simple-cubic Ising models:
Landau-Ginzburg-Wilson theory
{\em Phys. Rev.} B {\bf 30} 1362

\bibitem{Sokal-97}
Sokal A D 1997
Monte Carlo Methods in Statistical Mechanics: Foundations and New Algorithms
in Functional Integration: Basics and Applications (1996 Carg\`ese
School), ed C DeWitt-Morette, P Cartier and A Folacci (New York: Plenum)

\bibitem{Grassberger-95}
Grassberger P 1995
Damage spreading and critical exponents for model A Ising dynamics
{\em Physica} A {\bf 214} 547

(erratum) {\em Physica} A {\bf 217} 227

\end{thebibliography}
\end{document}